\documentclass[preprintnumbers,amsmath,amssymb,floatfix,11
pt,prd,superscriptaddress,nofootinbib]{revtex4}
\usepackage{graphicx}
\usepackage{epsfig}
\usepackage{bm}
\usepackage{amsfonts}

\input epsf

\begin{document}

\title{\Large Fractional Action Cosmology: Emergent,
\Large Logamediate,\\ Intermediate, Power Law Scenarios of the Universe and\\
\Large Generalized Second Law of Thermodynamics}

\author{Ujjal Debnath}
\email{ujjaldebnath@yahoo.com , ujjal@iucaa.ernet.in}
\affiliation{Department of Mathematics, Bengal Engineering and
Science University, Shibpur, Howrah-711 103, India.}

\author{Mubasher Jamil}
\email{mjamil@camp.nust.edu.pk , jamil.camp@gmail.com}
\affiliation{Center for Advanced Mathematics and Physics (CAMP),
National University of Sciences and Technology (NUST), H-12,
Islamabad, Pakistan.}

\author{Surajit Chattopadhyay}
\email{surajit_2008@yahoo.co.in,
surajit.chattopadhyay@pcmt-india.net} \affiliation{Department of
Computer Application (Mathematics Section), Pailan College of
Management and Technology, Bengal Pailan Park, Kolkata-700 104,
India.}

\begin{abstract}
\vspace*{1.5cm} \centerline{\bf Abstract} \vspace*{1cm}

In the framework of Fractional Action Cosmology (FAC), we study
the generalized second law of thermodynamics for the Friedmann
Universe enclosed by a boundary. We use the four well-known cosmic
horizons as boundaries namely, apparent horizon, future event
horizon, Hubble horizon and particle horizon. We construct the
generalized second law (GSL) using and without using the first law
of thermodynamics. To check the validity of GSL, we express the
law in the form of four different scale factors namely emergent,
logamediate, intermediate and power law. For Hubble, apparent and
particle horizons, the GSL holds for emergent and logamediate
expansions of the universe when we apply with and without using
first law. For intermediate scenario, the GSL is valid for Hubble,
apparent, particle horizons when we apply with and without first
law. Also for intermediate scenario, the GSL is valid for event
horizon when we apply first law but it breaks down without using
first law. But for power law expansion, the GSL may be valid for
some cases and breaks down otherwise.
\end{abstract}


\maketitle

\newpage

\section{Introduction}

From several cosmological observations \cite{perl}, it is now well
accepted in the scientific community that the observable Universe
is undergoing an accelerated expansion. This cosmic acceleration
is presumably driven some sort of a dense component having
negative pressure, named Dark Energy (DE)\cite{dark}. Note that,
the most powerful quantity of DE is its equation-of-state (EoS)
effectively defined as $w_{DE}=p_{DE}/\rho_{DE}$, where $p_{DE}$
and $\rho_{DE}$ are the pressure and energy density respectively.
There are several candidates of dark energy while the simplest one
is the cosmological constant \cite{pady}. However there are
problems with it, for instance, why its theoretical value based on
quantum field theory differs several orders of magnitude from the
empirical value \cite{sahni}. To model DE, there are other
approaches as well: modification of the Einstein-Hilbert
Lagrangian and replacing it with a general $f(R)$ function
\cite{nojiri}; modification of the Friedmann equation via adding
additional term in the density \cite{kath}; considering time
dependent and homogeneous scalar fields like phantom energy and
quintessence \cite{sami}, theories of extra-dimensions like
braneworld idea \cite{sa} or describing the acceleration via
choosing suitable scale factor rather then a dark fluid
\cite{deb}. We employ the last approach in this paper.

Here, we study the validity of a generalized second law (GSL) of
thermodynamics in the intermediate, logamediate, power law and
emergent scenarios of the universe in the framework of Fractional
Action Cosmology (FAC) discussed in the subsequent section.
According to the GSL, for our system, the sum of the entropy of
matter enclosed by the horizon and the entropy of the horizon must
not be a decreasing function of time \cite{setare}. A handful of
works are available where the GSL has been studied for various
candidates of dark energy in various interacting situations.
Izquierdo and Pavon \cite{pavon} explored the thermodynamic
consequences of a phantom-dominated universe and mentioned that
one must take into account that an ever accelerating universe has
a future event horizon (or cosmological horizon). In a recent
study, Setare and Sheykhi \cite{setare1} investigated RSII
braneworld filled with interacting viscous dark energy and dark
matter and concluded that in an accelerating universe with spatial
curvature, the apparent horizon is a physical boundary from the
thermodynamical point of view. Chattopadhyay and Debnath of
reference [9] studied the validity of the GSL in presence of
interacting tachyonic field and scalar (phantom) field as well as
in the presence of interaction between DBI-essence and other four
candidates of dark energy, namely the modified Chaplygin gas,
hessence, tachyonic field and new agegraphic dark energy. The said
authors found that in all cases, except for the phantom field
dominated universe, the GSL is breaking down. In studying the GSL,
we consider two possibilities \cite{gsl}: (1) in addition to the
entropy of all the matter, there is an entropy associated with the
horizon of the Universe, (2) local thermal equilibrium condition,
otherwise there will be a heat flow across the horizon which will
destroy the background FRW geometry. According to GSL, entropy of
everything (matter, radiation, dark energy) inside the horizon
added with the entropy of horizon must not decrease with time. In
the framework of Friedmann cosmology, this law can be validated
using two approaches: (1) employing the first law of
thermodynamics and determining the entropy relation on the
horizon, (2) using the well-known expressions of horizon entropy
and horizon temperature. Note that in the case of apparent
horizon, the two approaches are equivalent.

Organization of the rest of the paper is as follows: In section
II, we discuss the theoretical background of the fractional action
cosmological model and the way we consider interacting dark
energy. The emergent, logamediate and intermediate scenarios are
discussed in sections III, IV and V respectively. The GSL is
described in section VI. In this section we discuss the GSL using
as well as without using first law of thermodynamics. The GSL in
the said scenarios are discussed in the sections VII, VIII and IX
respectively. In section X we consider the power law form of the
scale factor and discuss the validity of the GSL. We conclude in
section XI.

\section{Fractional Action Cosmological Model}

Fractional action cosmology (FAC) is based on the formalism of the
fractional calculus. In this formalism, the order of
differentiation or integration is not an integer but a fractional
number. The fractional calculus is immensely useful in various
branches of mathematics, physics and engineering \cite{rami}. In
doing FAC, one can proceed in two different ways \cite{mark}: the
first one is quite easy as one has to replace the partial
derivatives in the Einstein field equations with the corresponding
fractional derivatives; the second technique involves deriving the
field equations and geodesic equations from a more fundamental
way, namely starting with the principle of least action and
replacing the usual integral with a fractional integral. This
later technique is more useful in giving extra features of the FAC
\cite{Rami}: Rami introduced the FAC by introducing the fractional
time integral,
\begin{equation}
S=-\frac{K}{2\Gamma(\xi)}\int\dot{x}^\mu\dot{x}^\nu
g_{\mu\nu}(x)(t-\tau)^{\xi-1}d\tau
\end{equation}
Here $\Gamma(\xi)=\int_0^\infty t^{\xi-1}e^{-t}dt$ is the Gamma
function, $0<\xi\leq1$, $0<\tau <t$, $K=$ constant and $\dot
x^\mu=\frac{dx^\mu}{d\tau}$.

For a FRW spacetime, the line element is given by
\begin{equation}
ds^{2} = - dt^{2} + a^{2}(t) \left[\frac{ dr^{2}}{1-k r^{2}} +
r^{2}(d\theta^{2}+\sin^{2}\theta d\phi^{2})\right]
\end{equation}
where $a(t)$ is the scale factor and $k ~(= 0, \pm 1)$ is the
curvature scalar.  The Einstein equations for the space-time given
by equation (2) are \cite{Rami}
\begin{eqnarray}
H^{2}+\frac{2(\xi-1)}{T_1}H+\frac{k}{a^{2}}&=&\frac{8\pi
G}{3}\rho\\
\dot{H}-\frac{(\xi-1)}{T_1}H-\frac{k}{a^{2}}&=& -4\pi G (\rho+p)
\end{eqnarray}
where $T_1=t-\tau$. Here $\rho$ and $p$ are the energy density and
pressure of the fluid.

\section{Emergent Scenario}

Here we consider the emergent scenario of the the Universe which
has the following properties: 1. it is almost static at the
infinite past ($t\rightarrow-\infty$) and isotropic, homogeneous
at large scales; 2. it is ever existing and there is no timelike
singularity; 3. the Universe is always large enough so that the
classical description of space-time is adequate; 4. the Universe
may contain exotic matter so that the energy conditions may be
violated; 5. the Universe is accelerating as suggested by recent
measurements of distances of high redshift type Ia supernovae.

To satisfy the above properties for emergent Universe, the scale
factor can be chosen as \cite{deb}
\begin{equation}
a(t) = a_{0}\left(\lambda+e^{\mu T_1}\right)^{n}
\end{equation}
where $a_{0},~\mu,~\lambda$ and $n$ are positive constants. (1) $a_0
> 0$ for the scale factor $a$ to be positive; (2) $\lambda > 0$, to
avoid any singularity at finite time (big-rip); (3) $a > 0$ or $n >
0$ for expanding model of the Universe; (4) $a < 0$ and $n < 0$
implies big bang singularity at $T_1 = -\infty$.

 So the
Hubble parameter and its derivatives are given by
\begin{equation}
H=\frac{n\mu e^{\mu T_1}}{\left(\lambda+e^{\mu T_1}\right)}~,~
\dot{H}=\frac{n\lambda\mu^{2}e^{\mu T_1}}{\left(\lambda+e^{\mu
T_1}\right)^{2}}~,~\ddot{H}=\frac{n\lambda\mu^{3}e^{\mu
T_1}(\lambda-e^{\mu T_1})}{\left(\lambda+e^{\mu T_1}\right)^{3}}
\end{equation}
Here $H$ and $\dot{H}$ are both positive, but $\ddot{H}$ changes
sign at $T_1=\frac{1}{\mu}~\text{log}\lambda$. Thus $H,~\dot{H}$ and
$\ddot{H}$ all tend to zero as $T_1\rightarrow -\infty$. On the
other hand as $T_1\rightarrow \infty$ the solution gives
asymptotically a de Sitter Universe.

Some recent works have considered the emergent scenario of the
universe in the framework of dark energy. Mukherjee et al
\cite{mukherjee} presented a general framework for an emergent
universe scenario and showed that emergent universe scenarios are
not isolated solutions and they may occur for different
combinations of radiation and matter. Campo et al \cite{campo}
studied the emergent universe model in the context of a
self-interacting Jordan-Brans-Dicke theory and showed that the
model presents a stable past eternal static solution which
eventually enters a phase where the stability of this solution is
broken leading to an inflationary period. In another study,
Debnath of reference [9] discussed the behaviour of different
stages of the evolution of the emergent universe considering that
the universe is filled with normal matter and a phantom field.
Paul et al \cite{paul} predicted the range of the permissible
values for the parameters associated with the constraints on
exotic matter needed for an emergent universe. In the present
work, we shall investigate the GSL in the emergent scenario for
the interacting dark energy stated earlier.

\section{Logamediate Scenario}

In this section we discuss the interacting dark energy under
consideration in the logamediate scenario. We consider a
particular form of logamediate scenario, where the form of the
scale factor $a(T_1)$ is defined as
\begin{equation}
a(T_1)=e^{A(\ln T_1)^{\alpha}}
\end{equation}
where $A \alpha>0$ and $\alpha>1$. When $\alpha=1$, this model
reduces to power-law form. Barrow and Nunes \cite{barrow}
considered this form of scale factor, where the cosmological scale
factor expands in the form expressed in the equation (12). This
form of scale factor has also been used in Khatua and Debnath
\cite{piyali} and in the first two works under reference [9]. The
logamediate form is motivated by considering a class of possible
cosmological solutions with indefinite expansion which result from
imposing weak general conditions on the cosmological model. Barrow
and Nunes \cite{barrow} found in their model that the
observational ranges of the parameters are as follows: $1.5\times
10^{-92}\le A \le 2.1\times 10^{-2}$ and $2\le \alpha \le 50$. The
Hubble parameter $H=\frac{\dot{a}}{a}$ and its derivative become,
\begin{equation}
H=\frac{A\alpha}{T_1}(\ln
T_1)^{\alpha-1}~~,~~\dot{H}=\frac{A\alpha}{T_1^{2}}(\ln
T_1)^{\alpha-2}(\alpha-1-\ln T_1).
\end{equation}

\section{Intermediate Scenario}

In the particular scenario of `intermediate' form, the expansion
scale factor $a(T_{1})$ of the Friedmann universe evolves as
\cite{barrow1}
\begin{equation}
a(T_1)=e^{B T_1^\beta}
\end{equation}
 where $B\beta>0$, $B>0$ and $0<\beta<1$. Here the expansion of Universe is faster
 than Power-Law
form, where the scale factor is given as, $a(T_1) = T_1^n$, where
$n>1$ is a constant. Also, the expansion of the Universe is slower
for Standard de-Sitter Scenario where $\beta = 1$.
 The Hubble parameter $H=\frac{\dot{a}}{a}$ and its derivative become,
\begin{equation}
H=B\beta T_1^{\beta-1}~,~~\dot{H}=B\beta(\beta-1) T_1^{\beta-2}.
\end{equation}

\section{Generalized Second Law of Thermodynamics: General Overview}

Here we extend the work of \cite{das} on thermodynamics in
fractional action cosmology. We denote the radius of cosmological
horizon by $R_{X}$. For Hubble, apparent, particle and event
horizon we replace $X$ by $H$, $A$, $P$ and $E$ respectively. The
corresponding radii are given by

\begin{equation}
R_{H}=\frac{1}{H}~;~~~R_{A}=\frac{1}{\sqrt{H^{2}+\frac{k}{a^{2}}}}~;~~~
R_{P}=a\int_{0}^{T_1}\frac{dT_1}{a}~;~~~R_{E}=a\int_{T_1}^{\infty}\frac{dT_1}{a}
\end{equation}

It can be easily obtained that

\begin{equation}
\dot{R}_{H}=-\frac{\dot{H}}{H^{2}}~;
~~\dot{R}_{A}=-HR_{A}^{3}\left(\dot{H}-\frac{k}{a^{2}}\right)~;
~~\dot{R}_{P}=HR_{P}+1~;~~\dot{R}_{E}=HR_{E}-1
\end{equation}

 To study the generalized second law (GSL) of thermodynamics through the
universe we deduce the expression for normal entropy using the
Gibb's equation of thermodynamics

\begin{equation}
T_{X}dS_{IX}=pdV_{X}+dE_{IX}
\end{equation}

where, $S_{IX}$ is the internal entropy within the horizon. Here
the expression for internal energy can be written as $E_{IX}=\rho
V_{X}$,  where the volume of the sphere is $V_{X}=\frac{4}{3}\pi
R_{X}^{3}$. Using equation (13) we obtain the rate of change of
internal energy as

\begin{equation}
\dot{S}_{IX}=\frac{4\pi
R_{X}^{2}}{T_{X}}(\rho+p)(\dot{R}_{X}-HR_{X})
\end{equation}

In the following, we shall find out the expressions of the rate of
change of total entropy using first law and without using first
law of thermodynamics.\\

\subsection{\large GSL using first law}

From the first law of thermodynamics, we have

\begin{equation}
T_{X}dS_{X}=4\pi R_{X}^{3}H (\rho+p)dT_1
\end{equation}

where, $T_{X}$ and $R_{X}$ are the temperature and radius of the
horizons under consideration in the equilibrium thermodynamics.\\

Using (15) we can get the time derivative of the entropy on the
horizon as

\begin{equation}
\dot{S}_{X}=\frac{4\pi R_{X}^{3}H}{T_{X}}(\rho+p).
\end{equation}

Adding equations (14) and (16) we get the time derivative of total
entropy as

\begin{equation}
\dot{S}_{X}+\dot{S}_{IX}=\frac{
R_{X}^{2}}{GT_{X}}\left(\frac{k}{a^{2}}+\frac{(\xi-1)}{T_1}H-\dot{H}\right)\dot{R}_{X}
\end{equation}

In order the GSL to be hold, we require
$\dot{S}_{X}+\dot{S}_{IX}\geq0$.\\

\subsection{\large GSL without using first law}

In this work, we shall also investigate the GSL without using the
first law of thermodynamics. The horizon entropy is $S_{X}=\frac{\pi
R_{X}^{2}}{G}$ and the temperature is $T_{X}=\frac{1}{2\pi R_{X}}$.
In this case, the time derivative of the entropy on the horizon is

\begin{equation}
\dot{S}_{X}=\frac{2\pi R_{X}\dot{R}_{X}}{G}
\end{equation}

Therefore, in this case the time derivative of the total entropy is

\begin{equation}
\dot{S}_{X}+\dot{S}_{IX}=\frac{2\pi
R_{X}}{G}\left[R_{X}^{2}\left(\frac{k}{a^{2}}+\frac{(\xi-1)}{T_1}H
-\dot{H}\right)\left(\dot{R}_{X}-HR_{X}\right)+ \dot{R}_{X}\right].
\end{equation}
In the following sections, we shall investigate the nature of the
equations (17) and (19) i.e., validity of GSL in four scenarios,
namely, \emph{emergent}, \emph{logamediate}, \emph{intermediate}
and \emph{power law} scenarios.

\section{\normalsize\bf{GSL in Emergent Scenario}}

For emergent scenario, the Hubble, apparent, particle and event
horizon radii can be calculated as
\begin{equation}
\begin{array}{c}
R_{H}=\frac{(\lambda+e^{\mu T_1})}{n\mu e^{\mu
T_1}}~;~~R_{A}=\left[\frac{n^{2}\mu^{2}e^{2\mu T_1}}
{\left(\lambda+e^{\mu T_1}\right)^{2}}+\frac{k}{a_{0}^{2}\left(\lambda+e^{\mu T_1
}\right)^{2n} } \right]^{-\frac{1}{2}}~;~~ \\\\
R_{P}=\frac{1}{n\mu}~\left(\lambda+e^{\mu
T_1}\right)^{n}\left\{_{2}F_{1}[n,n,n+1,-\lambda] -e^{-n\mu T_1}~
_{2}F_{1}[n,n,n+1,-\lambda e^{-\mu T_1}]\right\}~;~~\\\\
R_{E}=\frac{1}{n\mu}~\left(\lambda+e^{\mu T_1}\right)^{n}e^{-n\mu T_1}
~_{2}F_{1}[n,n,n+1,-\lambda e^{-\mu T_1}] \\
\end{array}
\end{equation}

Using (17), (19) and (20) we get the time derivatives of the total
entropies to investigate the validity of the GSL in various
horizons using and without using first law.\\

\subsection{\bf GSL in the \emph{emergent} scenario using first law}

Here we consider the GSL in the \emph{emergent} scenario using the
first law of thermodynamics. Using (17) and (20) we get the
time derivative of total entropies as follows:\\
\begin{itemize}
    \item For Hubble horizon
    \begin{equation}
    \dot{S}_{H}+\dot{S}_{IH}=\frac{\lambda e^{-3\mu T_1}
    \left[-kT_1(\lambda+e^{\mu T_1})^{2-2n}+a_{0}^{2}n\mu e^{\mu T_1}
    \left(\lambda\mu T_1-(\xi-1)(\lambda+e^{\mu T_1})\right)\right]}
    {a_{0}^{2}n^{3}\mu^{2}T_1GT_{H}}
    \end{equation}
    \item For apparent horizon
    \begin{equation}
     \dot{S}_{A}+\dot{S}_{IA}=\frac{a_{0}n\mu e^{\mu T_1}
     \left[k(\lambda+e^{\mu T_1})^{2-2n}-a_{0}^{2}n\lambda\mu^{2} e^{\mu T_1}
    \right]\left[kt(\lambda+e^{\mu T_1})^{2-2n}+a_{0}^{2}n\mu e^{\mu T_1}
    \left((\xi-1)(\lambda+e^{\mu T_1})-\lambda\mu T_1\right)\right]}{GT_{A}
    T_1\left[k(\lambda+e^{\mu T_1})^{2-2n}+a_{0}^{2}n^{2}\mu^{2} e^{2\mu T_1}
    \right]^{\frac{5}{2}}}~~~~~~~~~~~~~~~~~~~~~\\
    \end{equation}
    \item For particle horizon
    \begin{eqnarray*}
    \dot{S}_{P}+\dot{S}_{IP}=\frac{(\lambda+e^{\mu T_1})^{3n-3}}{a_{0}^{2}
    n^{2}\mu^{2}T_1GT_{P}}~\left[kt(\lambda+e^{\mu T_1})^{2-2n}+a_{0}^{2}n\mu e^{\mu T_1}
    \left((\xi-1)(\lambda+e^{\mu T_1})-\lambda\mu
    T_1\right)\right]
      \end{eqnarray*}
      \begin{eqnarray*}
   \times \left[(\lambda+e^{\mu T_1})^{1-n}+e^{\mu T_1} \left\{_{2}
   F_{1}[n,n,n+1,-\lambda] -e^{-n\mu T_1}~
  _{2}F_{1}[n,n,n+1,-\lambda e^{-\mu T_1}]\right\}\right]
      \end{eqnarray*}
       \begin{equation}
     \times \left\{_{2}F_{1}[n,n,n+1,-\lambda] -e^{-n\mu T_1}~
  _{2}F_{1}[n,n,n+1,-\lambda e^{-\mu T_1}]\right\}^{2}\\
      \end{equation}
    \item For event horizon
    \begin{eqnarray*}
    \dot{S}_{E}+\dot{S}_{IE}=\frac{(\lambda+e^{\mu T_1})^{3n-3}}{a_{0}^{2}
    n^{2}\mu^{2}T_1GT_{E}}~\left[kT_1(\lambda+e^{\mu T_1})^{2-2n}+a
    _{0}^{2}n\mu e^{\mu T_1}
    \left((\xi-1)(\lambda+e^{\mu T_1})-\lambda\mu
    T_1\right)\right]
      \end{eqnarray*}
      \begin{equation}
     \times \left[-(\lambda+e^{\mu T_1})^{1-n}+e^{\mu T_1}~
  _{2}F_{1}[n,n,n+1,-\lambda e^{-\mu T_1}]\right] \left\{e^{-n\mu T_1}~
  _{2}F_{1}[n,n,n+1,-\lambda e^{-\mu T_1}]\right\}^{2}\\
      \end{equation}
\end{itemize}

\begin{figure}
\includegraphics[height=2in]{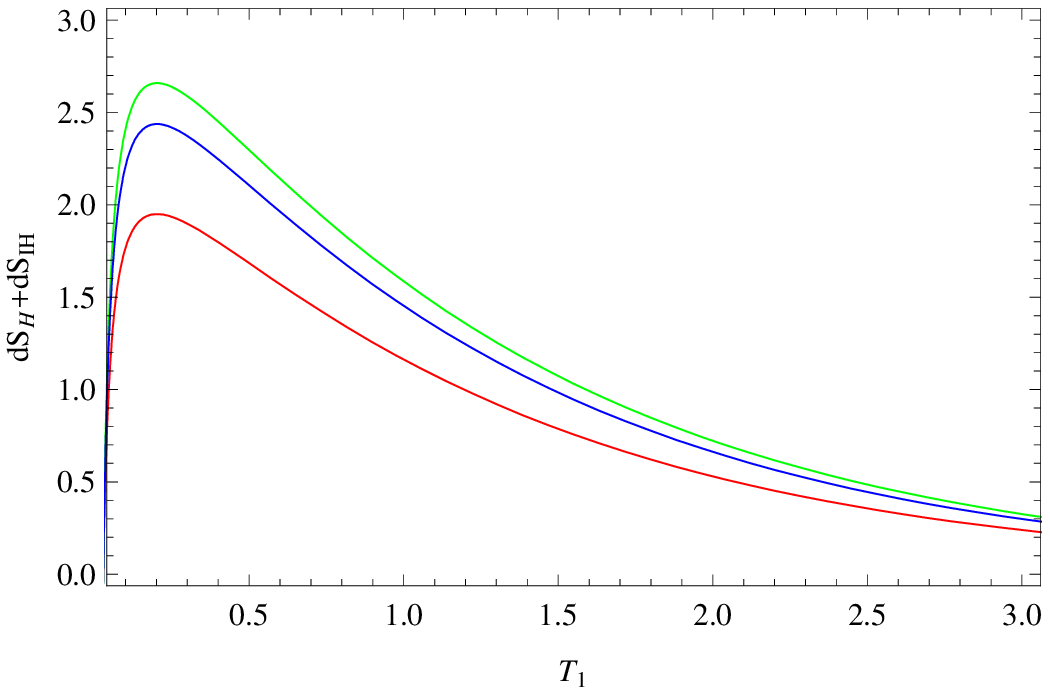}~~~~
\includegraphics[height=2in]{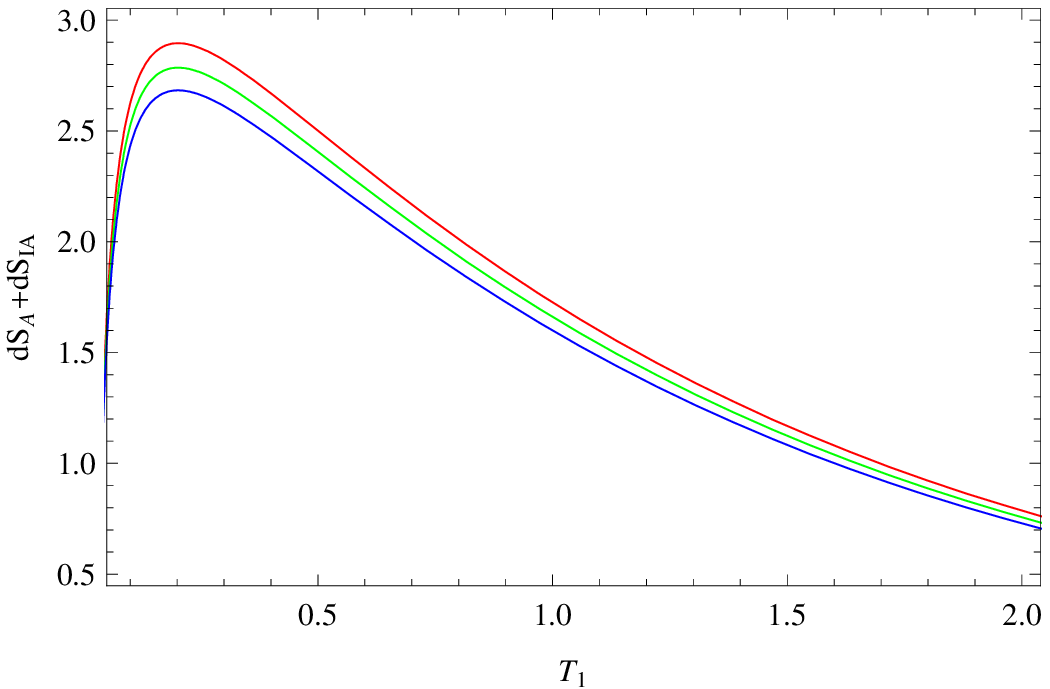}\\
\vspace{1mm} ~~~~~~~Fig.1~~~~~~~~~~~~~~~~~~~~~~~~~~~~~~~~~~~
~~~~~~~~~~~~~~~~~~~~~~~~~~~~~~~~~~~~~~Fig.2~~~\\

\vspace{6mm}

\includegraphics[height=2in]{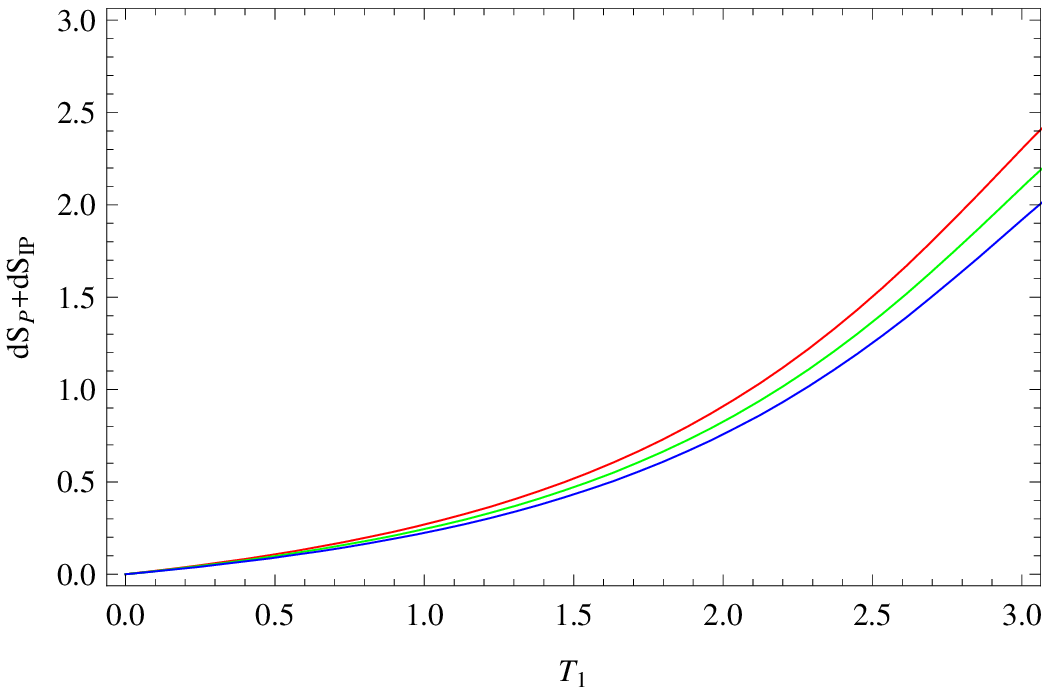}~~~~
\includegraphics[height=2in]{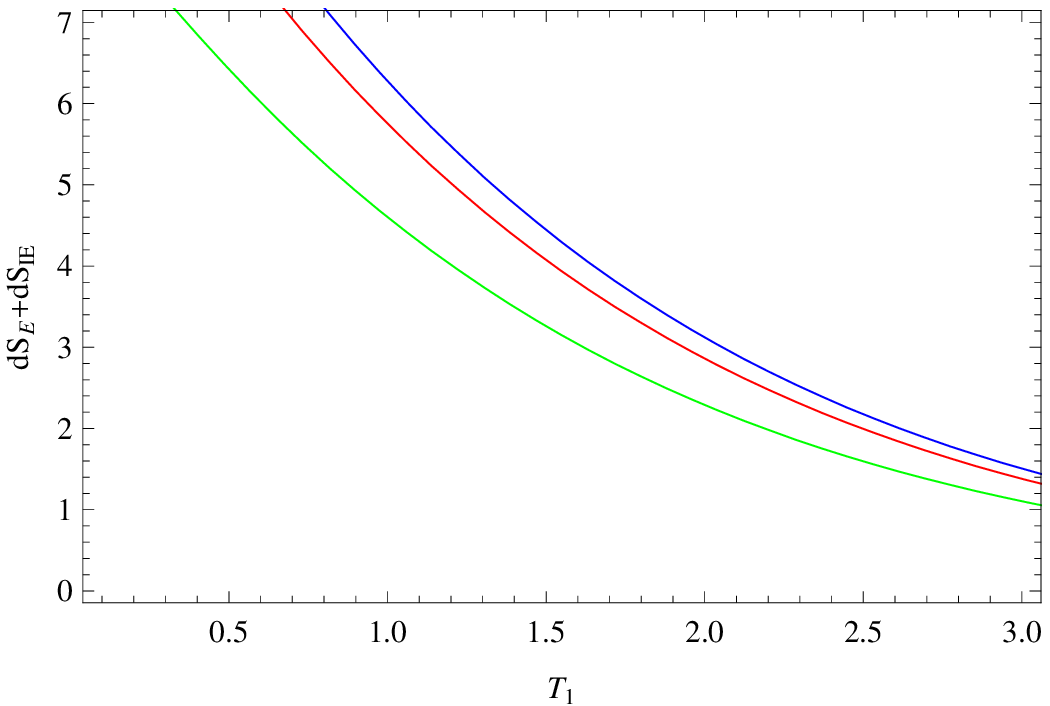}\\
\vspace{1mm} ~~~~~~~Fig.3~~~~~~~~~~~~~~~~~~~~~~~~~~~~~~~~~~~~~~~~
~~~~~~~~~~~~~~~~~~~~~~~~~~~~~~~~~Fig.4~~~\\

\vspace{6mm} Figs. 1, 2, 3 and 4 show the time derivatives of the
total entropy for Hubble horizon $R_{H}$, apparent horizon $R_{A}$,
particle horizon $R_{P}$ and event horizon $R_{E}$ respectively
\textbf{using first law of thermodynamics} in the \emph{emergent}
scenario. The red, green and blue lines represent the
$dS_{X}+dS_{IX}$ for $k=-1,~1$ and $0$
respectively. We have chosen $\xi=0.3$.\\

 \vspace{6mm}

 \end{figure}

It is easily observed from the figures 1 to 4, that the total
entropy increases only in the case of particle horizon. Also note
that the maximum rate of increase in entropy appears in the case
of spatially hyperbolic Universe while a flat Universe has the
lowest rate of increase in entropy. The GSL is valid for all
horizons.

\begin{figure}
 \includegraphics[height=2in]{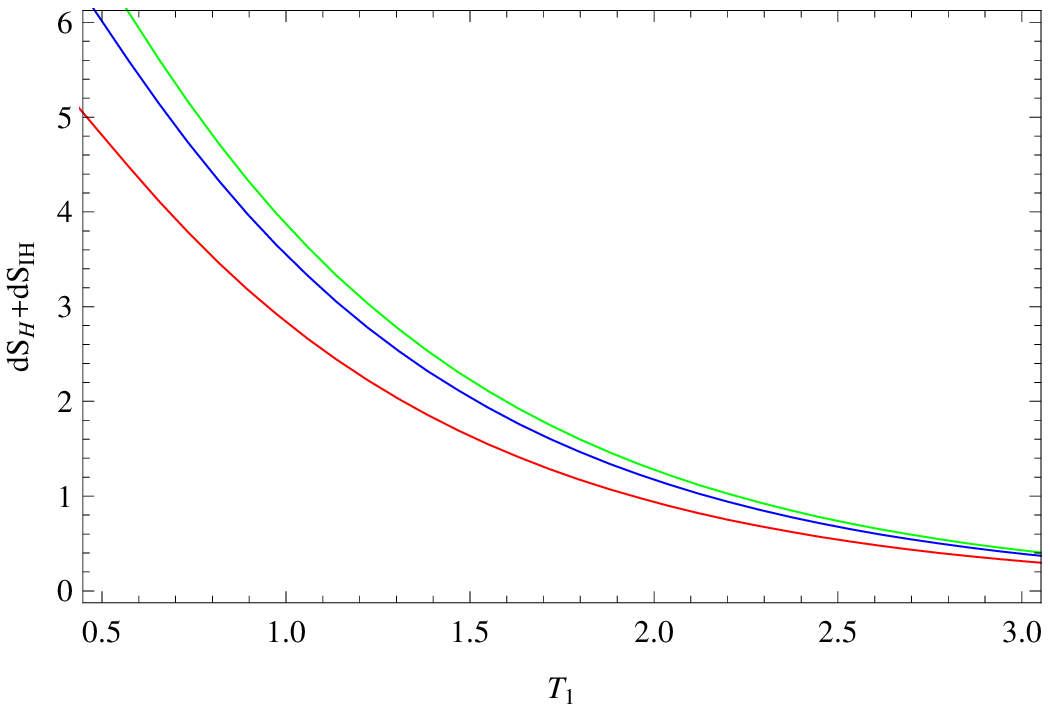}~~~~
\includegraphics[height=2in]{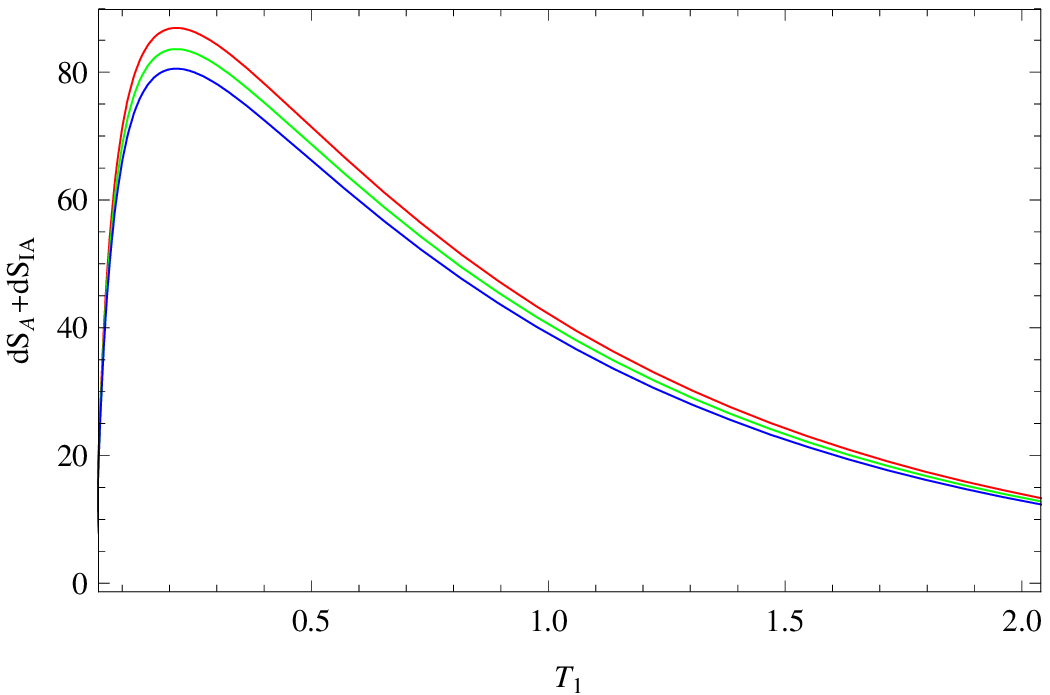}\\
\vspace{1mm} ~~~~~~~Fig.5~~~~~~~~~~~~~~~~~~~~~~~~~~~~~~~~~~~~~~~~~~~~
~~~~~~~~~~~~~~~~~~~~~~~~~~~~~Fig.6~~~\\

\vspace{6mm}
\includegraphics[height=2in]{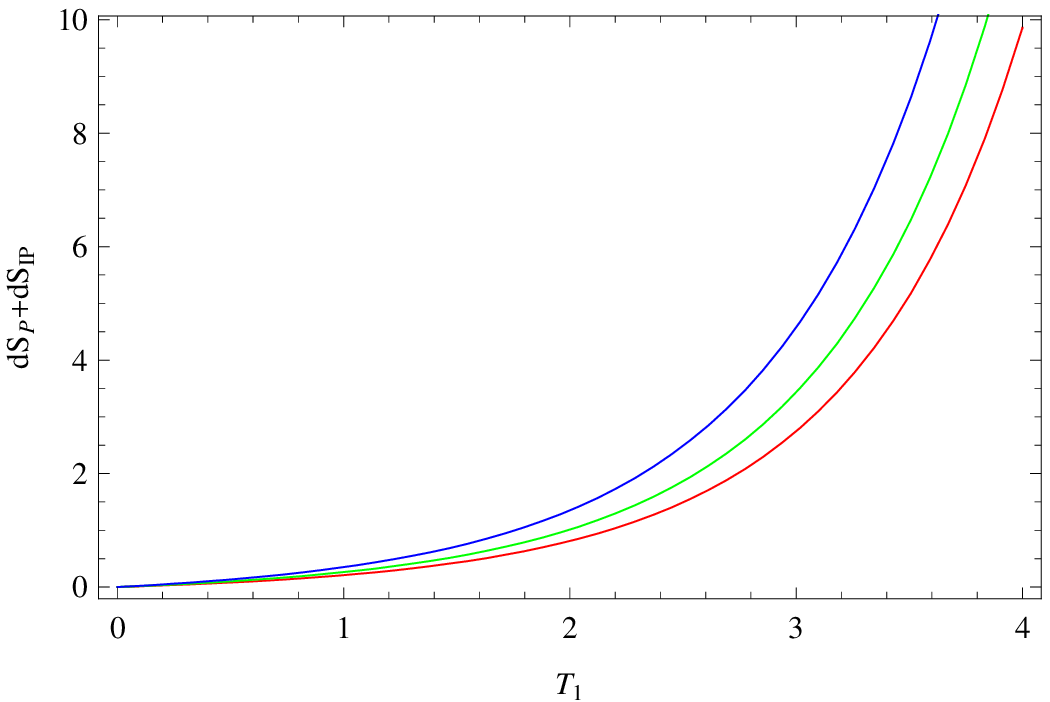}~~~~
\includegraphics[height=2in]{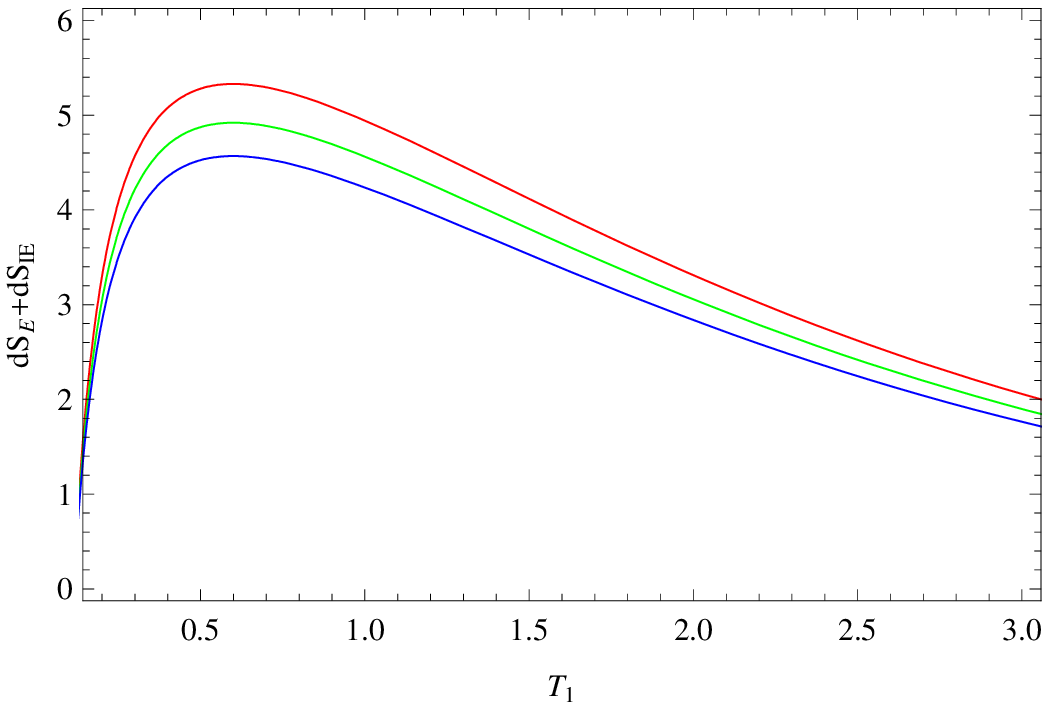}\\
\vspace{1mm} ~~~~~~~Fig.7~~~~~~~~~~~~~~~~~~~~~~~~~~~~~~~~~~~~~~~~
~~~~~~~~~~~~~~~~~~~~~~~~~~~~~~~~~Fig.8~~~\\

\vspace{6mm} Figs. 5, 6, 7 and 8 show the time derivatives of the
total entropy for Hubble horizon $R_{H}$, apparent horizon $R_{A}$,
particle horizon $R_{P}$ and event horizon $R_{E}$ respectively
\textbf{without using first law of thermodynamics} in the
\emph{emergent} scenario. The red, green and blue lines represent
the $dS_{X}+dS_{IX}$ for $k=-1,~1$ and $0$
respectively. We have chosen $\xi=0.3$.\\

 \vspace{3mm}

 \end{figure}

\subsection{\bf GSL in the \emph{emergent} scenario without using
first law}

Without using the first law, the time derivative of the total
entropies in the \emph{emergent} scenario come out as

\begin{itemize}
    \item For Hubble horizon
    \begin{eqnarray*}
    \dot{S}_{H}+\dot{S}_{IH}=\frac{2\pi e^{-4\mu T_1}(\lambda+e^{\mu T_1})}
    {a_{0}^{2}n^{4}\mu^{3}T_1G}
    \left[-kT_1(\lambda+ne^{\mu T_1})(\lambda+e^{\mu T_1})^{2-2n}
    \right.
    \end{eqnarray*}
    \begin{equation}
    \left. +a_{0}^{2}n\mu e^{\mu T_1}\left\{\lambda^{2}\mu T_1-(\xi-1)\left(ne^{2\mu T_1}
    +\lambda(n+1)e^{\mu T_1}+\lambda^{2}\right)\right\}\right]
    \end{equation}
    \item For apparent horizon
    \begin{eqnarray*}
     \dot{S}_{A}+\dot{S}_{IA}=\frac{2\pi\mu na_{0}^{2} e^{\mu T_1}(\lambda+e^{\mu T_1
     })}{Gt} \left[k^{2}T_1(\lambda+e^{\mu
     T_1})^{4-4n}-2knT_1\lambda\mu^{2}a_{0}^{2}(\lambda+e^{\mu
     T_1})^{2-2n}\right.
    \end{eqnarray*}
    \begin{equation}
    \left. +a_{0}^{4}n^{2}\mu^{3} e^{2\mu T_1}\left\{\lambda^{2}\mu T_1-
    (\xi-1)\left(ne^{2\mu T_1}+\lambda(n+1)e^{\mu T_1}
    +\lambda^{2}\right)\right\}\right]
    \end{equation}
    \item For particle horizon
    \begin{eqnarray*}
     \dot{S}_{P}+\dot{S}_{IP}=\frac{2\pi(\lambda+e^{\mu T_1})^{2n}}{n\mu G}
     ~\left\{_{2}F_{1}[n,n,n+1,-\lambda] -e^{-n\mu T_1}~
  _{2}F_{1}[n,n,n+1,-\lambda e^{-\mu T_1}]\right\}
    \end{eqnarray*}
    \begin{eqnarray*}
    \times\left[\frac{(\lambda+e^{\mu T_1})^{n-2}}{a_{0}^{2}n^{2}\mu^{2}T_1}
     \left\{kT_1(\lambda+e^{\mu T_1})^{2-2n}+a_{0}^{2}n\mu e^{\mu T_1}
    \left((\xi-1)(\lambda+e^{\mu T_1})-\lambda\mu
    T_1\right)\right\}  \right.
    \end{eqnarray*}
    \begin{eqnarray*}
    \times\left\{_{2}F_{1}[n,n,n+1,-\lambda] -e^{-n\mu T_1}~
    _{2}F_{1}[n,n,n+1,-\lambda e^{-\mu T_1}]\right\}^{2}
    \end{eqnarray*}
    \begin{eqnarray*}
   \left.  +\frac{e^{\mu T_1}}{(\lambda+e^{\mu T_1})}\left\{_{2}F_{1}[n,n,n+1,-\lambda]
    -e^{-n\mu T_1}~
    _{2}F_{1}[n,n,n+1,-\lambda e^{-\mu T_{1}}]\right\}+(\lambda+e^{\mu
    T_{1}})^{-n}\right]
    \end{eqnarray*}
    \item For event horizon
    \begin{eqnarray*}
     \dot{S}_{E}+\dot{S}_{IE}=\frac{2\pi e^{-n\mu T_1}(\lambda+e^{\mu T_1})^{2n-2}}
     {a_{0}^{2}n^{3}\mu^{3}T_1 G}~
     \left\{~_{2}F_{1}[n,n,n+1,-\lambda e^{-\mu T_1}]\right\}
    \end{eqnarray*}
      \begin{eqnarray*}
    \times \left[-a_{0}^{2}n^{2}\mu^{2}T_1(\lambda+e^{\mu T_1})^{2-n}+ a_{0}^{2}
    n^{2}\mu^{2}T_1 e^{(n-1)\mu T_1} (\lambda+e^{\mu T_1})
      ~_{2}F_{1}[n,n,n+1,-\lambda e^{-\mu T_1}] \right.
      \end{eqnarray*}
    \begin{equation}
   \left. + \left\{-kT_1(\lambda+e^{\mu T_1})^{2-2n}+a_{0}^{2}n\mu e^{\mu T_1}
    \left(\lambda\mu T_1-(\xi-1)(\lambda+e^{\mu T_1})\right)\right\}
    \left\{_{2}F_{1}[n,n,n+1,-\lambda e^{-\mu
    T_1}]\right\}^{2}\right]
    \end{equation}
    \end{itemize}

From figures 5 to 8, we see that the total entropy increases only
in the case of particle horizon. In the emergent scenario, the
maximum rate of increase in entropy appears in the case of
spatially flat Universe while a hyperbolic curved Universe has the
lowest rate of increase in entropy. The GSL is valid for all
horizons.

\section{\normalsize\bf{GSL in Logamediate Scenario}}

We obtain the radii of Hubble, apparent, particle and event
horizons in logamediate expansion as

\begin{equation}
\begin{array}{c}
 R_{H}=\frac{T_1(\ln T_1)^{1-\alpha}}{A\alpha}~;~~R_{A}=\frac{1}{\sqrt{e^{-2A(\ln T_1)^{\alpha}}k+\frac{A^{2}\alpha^{2}(\ln T_1)^{-2(1-\alpha)}}{T_{1}^{2}}}}~;~~ \\\\
  R_{P}=exp(A(\ln T_1)^{\alpha})\int_{0}^{T_1}\frac{dT_{1}}{exp(A(\ln T_1)^{\alpha})}~;~~R_{E}=exp(A(\ln T_1)^{\alpha})\int_{T_1}^{\infty}\frac{dT_{1}}{exp(A(\ln T_1)^{\alpha})} \\
\end{array}
\end{equation}

Using (17), (19) and (28) we get the time derivatives of the total
entropies to investigate the validity of the GSL in various
horizons using and without using first law.\\

\subsection{\bf GSL in the \emph{logamediate} scenario using first law}

Here we consider the GSL in the \emph{logamediate} scenario using
the first law of thermodynamics. The
time derivative of total entropies as follows:\\
\begin{itemize}
    \item For Hubble horizon
    \begin{equation}
    \dot{S}_{H}+\dot{S}_{IH}=\frac{e^{-2A(\ln T_1)^{\alpha}}(\alpha-1-\ln T_1)(\ln T_1)^{-3\alpha}\left(-kT_{1}^{2}(\ln T_1)^{2}
    +A\alpha e^{2A(\ln T_1)^{\alpha}}(\ln T_1)^{\alpha}(\alpha-1-\xi\ln T_1)   \right)}{A^{3}\alpha^{3}GT_{H}}
    \end{equation}
    \item For apparent horizon
    \begin{eqnarray*}
     \dot{S}_{A}+\dot{S}_{IA}=~~~~~~~~~~~~~~~~~~~~~~~~~~~~~~~~~~~~~~~~~~~~~~~~~~~~~~~~~~~~~~~~~~~~~~~~~~~~~~~~~~~~~~~~~~~~~~~~~~~~~~~
    \end{eqnarray*}
    \begin{equation}
     \frac{A\alpha e^{A(\ln T_1)^{\alpha}}(\ln T_1)^{\alpha}\left(kT_{1}^{2}(\ln T_1)^{2}
     -Ae^{2A(\ln T_1)^{\alpha}}\alpha(\alpha-1-\ln T_1)(\ln T_1)^{\alpha}\right)\left(kT_{1}^{2}(\ln T_1)^{2}
     -Ae^{2A(\ln T_1)^{\alpha}}\alpha(\alpha-1-\xi\ln T_1)(\ln T_1)^{\alpha}\right)}{GT_{A}\left(kT_{1}^{2}(\ln T_1)^{2}
     +A^{2}e^{2A(\ln T_1)^{\alpha}}\alpha^{2}(\ln
     T_1)^{2\alpha}\right)^{5/2}}  ~~~~~~~~~~~~~\\
    \end{equation}
    \item For particle horizon
    \begin{equation}
    \dot{S}_{P}+\dot{S}_{IP}=\frac{e^{-2A(\ln T_1)^{\alpha}} R_{P}^{2}\left(T_1\ln T_1+A \alpha (\ln T_1)^{\alpha}R_{P}\right)
    (kT_{1}^{2}(\ln T_1)^{2}-Ae^{2A(\ln T_1)^{\alpha}}\alpha(\alpha-1-\xi\ln T_1)(\ln T_1)^{\alpha})}{T_{1}^{3} (\ln T_1)^{3}GT_{P}}\\
      \end{equation}
    \item For event horizon
    \begin{equation}
    \dot{S}_{E}+\dot{S}_{IE}=\frac{e^{-2A(\ln T_1)^{\alpha}} R_{E}^{2}\left(-T_1\ln T_1+A \alpha (\ln T_1)^{\alpha}R_{E}\right)
    (kT_{1}^{2}(\ln T_1)^{2}-Ae^{2A(\ln T_1)^{\alpha}}\alpha(\alpha-1-\xi\ln T_1)(\ln T_1)^{\alpha})}{T_{1}^{3} (\ln T_1)^{3}GT_{E}}\\
      \end{equation}
\end{itemize}

\begin{figure}
\includegraphics[height=2in]{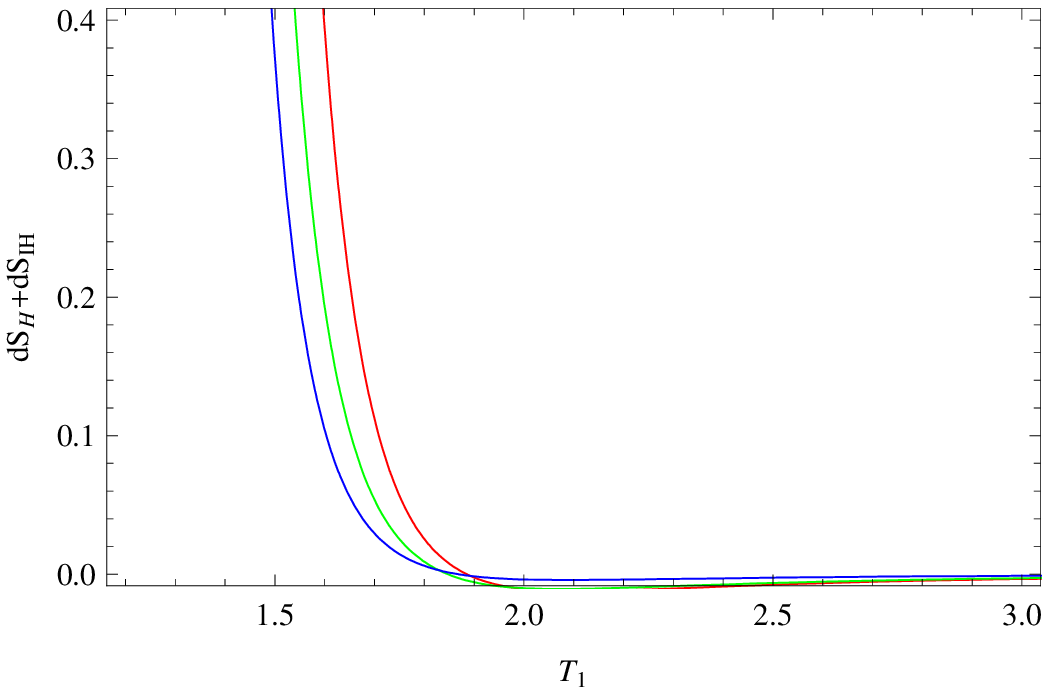}~~~~
\includegraphics[height=2in]{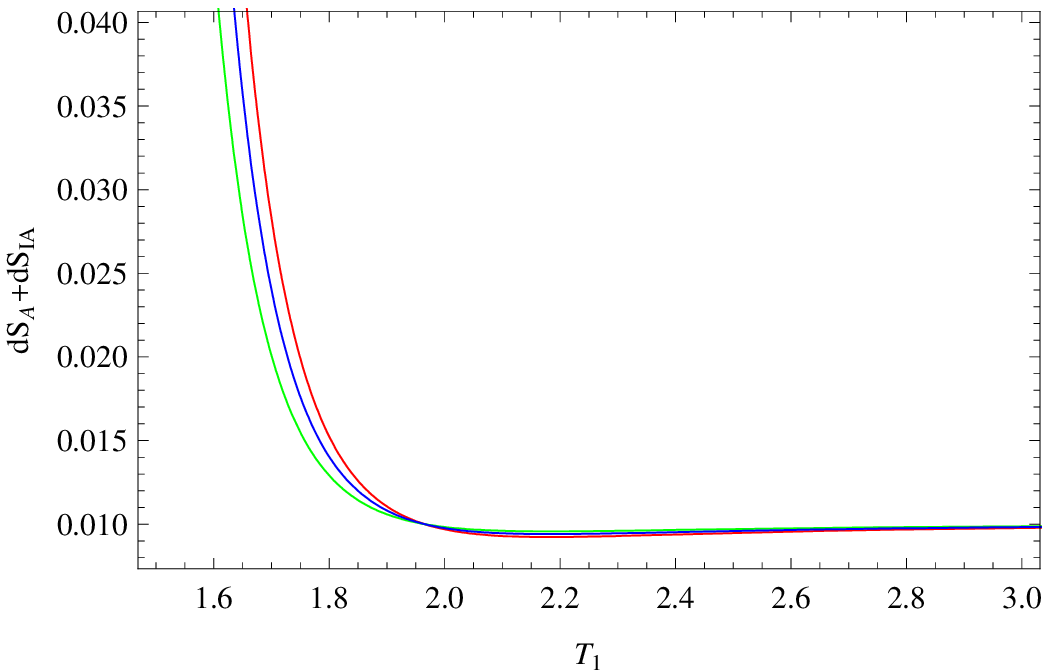}\\
\vspace{1mm} ~~~~~~~Fig.9~~~~~~~~~~~~~~~~~~~~~~~~~~~~~~~~~~~~
~~~~~~~~~~~~~~~~~~~~~~~~~~~~~~~~~~~~~Fig.10~~~\\

\vspace{6mm}

\includegraphics[height=2in]{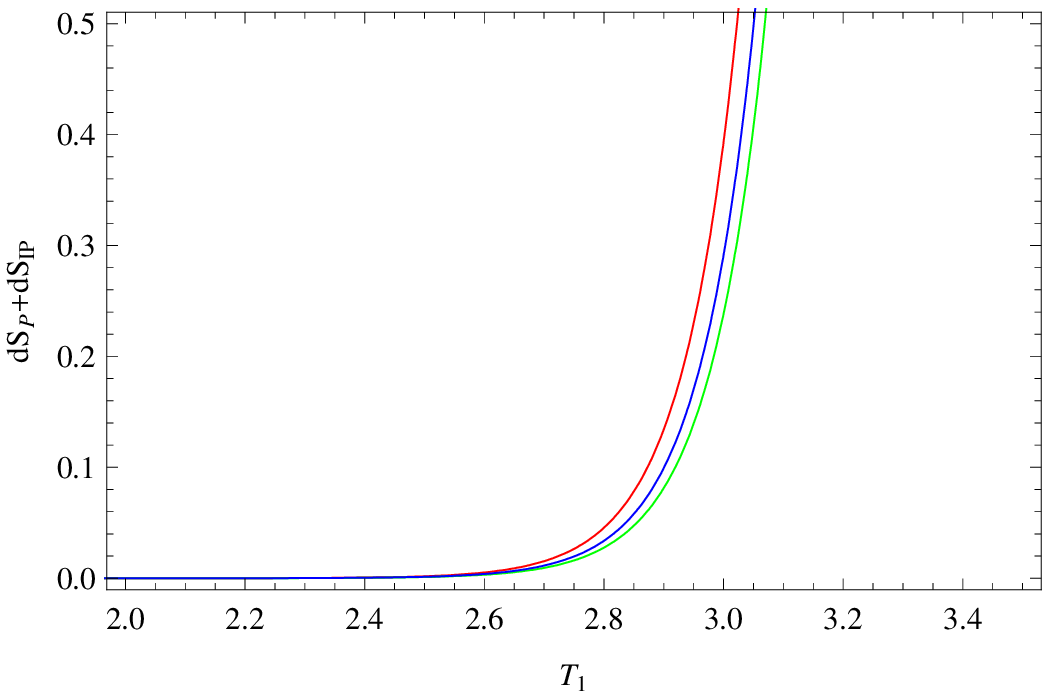}~~~~
\includegraphics[height=2in]{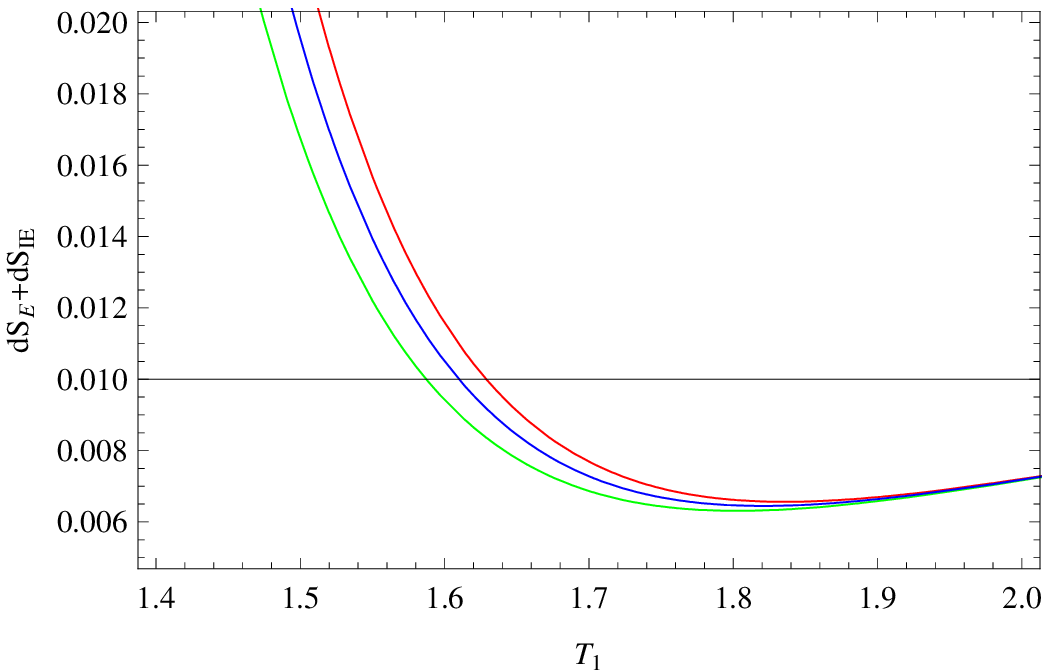}\\
\vspace{1mm} ~~~~~~~Fig.11~~~~~~~~~~~~~~~~~~~~~~~~~~~~~~~~~~~~~~~~
~~~~~~~~~~~~~~~~~~~~~~~~~~~~~~~~~Fig.12~~~\\

\vspace{6mm} Figs. 9, 10, 11 and 12 show the time derivatives of the
total entropy for Hubble horizon $R_{H}$, apparent horizon $R_{A}$,
particle horizon $R_{P}$ and event horizon $R_{E}$ respectively
\textbf{using first law of thermodynamics} in the \emph{logamediate}
scenario. The red, green and blue lines represent the
$dS_{X}+dS_{IX}$ for $k=-1,~1$ and $0$
respectively. We have chosen  $\xi=0.3$.\\

 \vspace{6mm}

 \end{figure}

From figures 9 to 12, we see that the total entropy increases only
in the case of particle horizon. In other words, the GSL holds for
all horizons in the Logamediate scenario. In figs.9, the total
change in entropy goes to zero in a finite time while in fig.10,
this rate remains constant but never tends to zero.

\subsection{\bf GSL in the \emph{logamediate} scenario without using
first law}

Without using the first law, the time derivative of the total
entropies in the \emph{logamediate} scenario come out as

\begin{itemize}
    \item For Hubble horizon
    \begin{eqnarray*}
    \dot{S}_{H}+\dot{S}_{IH}=~~~~~~~~~~~~~~~~~~~~~~~~~~~~~~~~~~~~~~~~~~~~~~~~~~~~~~~~~~~~~~~~~~~~~~~~~~~~~~~~~~~~~~~~~~~~~~~~~~~~~~~~~
    \end{eqnarray*}
    \begin{eqnarray*}
    \frac{2\pi T_1}{A^{4}\alpha^{4}G}~e^{-2A(\ln T_1)^{\alpha}} (\ln T_1)
    ^{1-4\alpha}\left[kT_1^{2}(\ln T_1)^{2}
    (1-\alpha+\ln T_1)+A\alpha (\ln T_1)^{\alpha}(-kT_1^{2}(\ln T_1)^{2}+e^{2A(\ln T_1)^{\alpha}}(1-\alpha+\ln
    T_1)^{2})\right.
    \end{eqnarray*}
    \begin{equation}
    \left.
    +(\xi-1)A\alpha e^{2A(\ln T_1)^{\alpha}}(\ln T_1)^{1+\alpha}
     (A\alpha(\ln T_1)^{\alpha}-1+\alpha-\ln T_1 ) \right]
    \end{equation}
    \item For apparent horizon
    \begin{eqnarray*}
     \dot{S}_{A}+\dot{S}_{IA}=\frac{2\pi T_1\alpha A}{G}~e^{2A(\ln T_1)^{\alpha}} (\ln T_1)^{1+\alpha}\left[\left(kT_1^{2}(\ln T_1)^{2}
     -Ae^{2A(\ln T_1)^{\alpha}}\alpha(\alpha-1-\ln T_1)(\ln
     T_1)^{\alpha}\right)^{2}\right.~~~~~~~~~~~~~~~~~~~~~~~
    \end{eqnarray*}
    \begin{equation}
    \left.  -(\xi-1)A^{2}\alpha^{2}e^{4A(\ln T_1)^{\alpha}}(\ln T_1)^{1+2\alpha}
     (A\alpha(\ln T_1)^{\alpha}-1+\alpha-\ln T_1 )\right] \left(kT_1^{2}(\ln T_1)^{2}+A^{2}\alpha^{2}e^{2A(\ln T_1)^{\alpha}}(\ln
     T_1)^{2\alpha}\right)^{-3}
    \\
    \end{equation}
    \item For particle horizon
    \begin{equation}
    \dot{S}_{P}+\dot{S}_{IP}=\frac{2\pi R_{P}^{3}}{G}\left(ke^{-2A(\ln T_1)^{\alpha}}-\frac{A\alpha(\alpha-1)(\ln T_1)^{-2+\alpha}}{T_1^{2}}
    +\frac{A\alpha \xi (\ln T_1)^{\alpha-1}}{T_1^{2}}\right)+\frac{2\pi R_{P}}{G}\left(\frac{A\alpha R_{P}(\ln T_1)^{\alpha-1}}{T_1}+1\right)
    \end{equation}
    \item For event horizon
    \begin{equation}
      \dot{S}_{E}+\dot{S}_{IE}=-\frac{2\pi R_{E}^{3}}{G}\left(ke^{-2A(\ln T_1)^{\alpha}}-\frac{A\alpha(\alpha-1)(\ln T_1)^{-2+\alpha}}{T_1^{2}}
      +\frac{A\alpha \xi (\ln T_1)^{\alpha-1}}{T_1^{2}}\right)+\frac{2\pi R_{E}}{G}\left(\frac{A\alpha R_{E}(\ln T_1)^{\alpha-1}}{T_1}-1\right)
    \end{equation}
    \end{itemize}

\begin{figure}
 \includegraphics[height=2in]{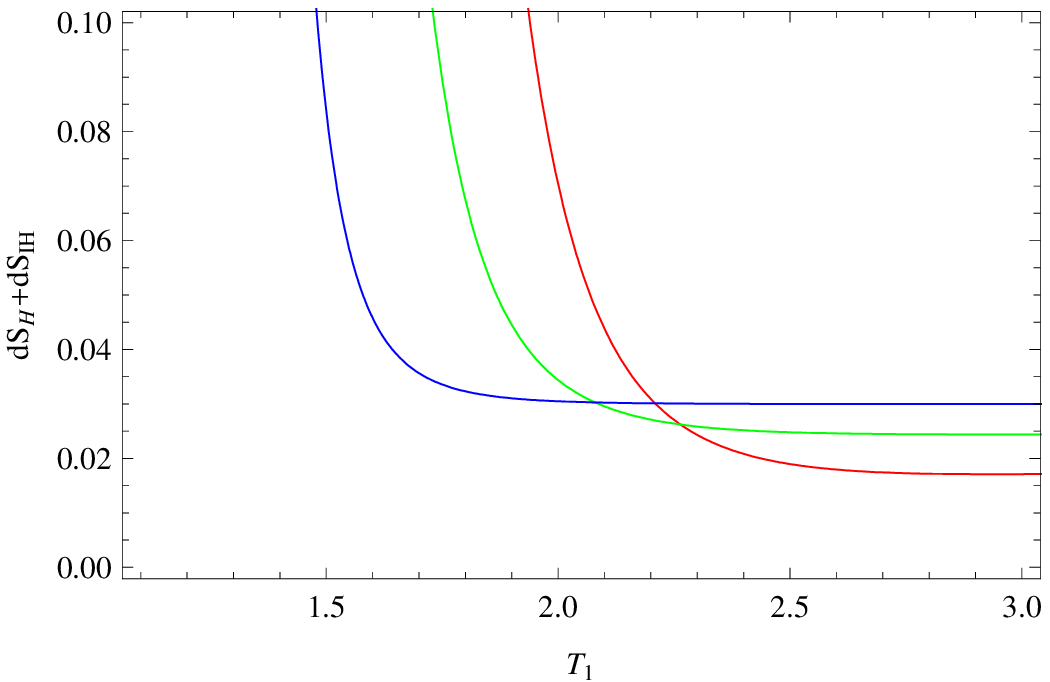}~~~~
\includegraphics[height=2in]{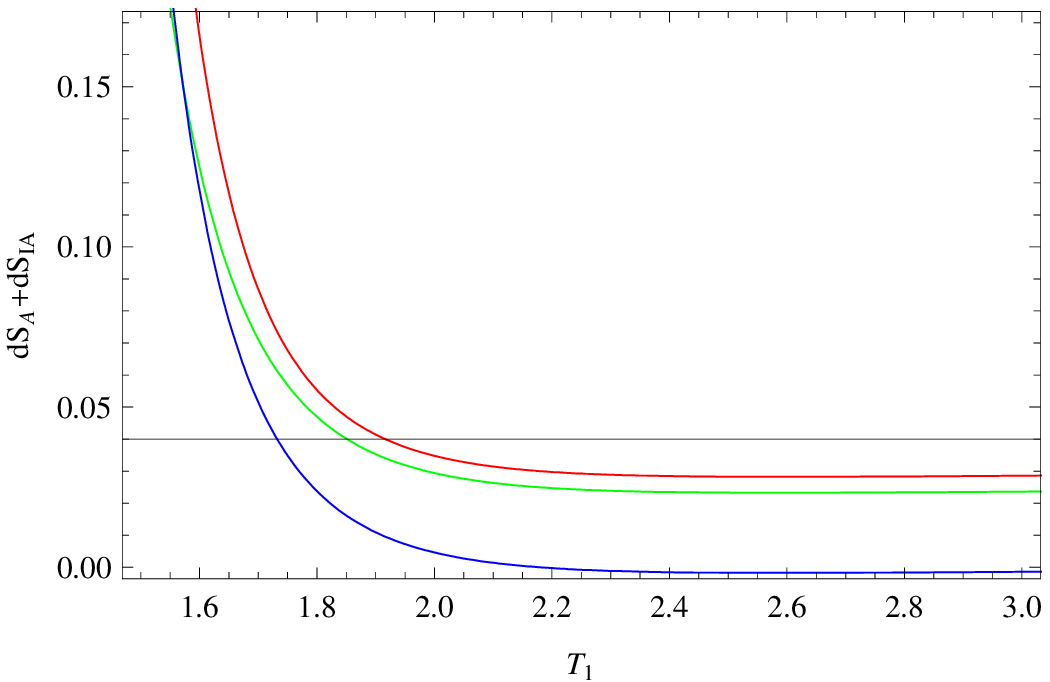}\\
\vspace{1mm} ~~~~~~~Fig.13~~~~~~~~~~~~~~~~~~~~~~~~~~~~
~~~~~~~~~~~~~~~~~~~~~~~~~~~~~~~~~~~~~~~~~~~~~Fig.14~~~\\

\vspace{6mm}
\includegraphics[height=2in]{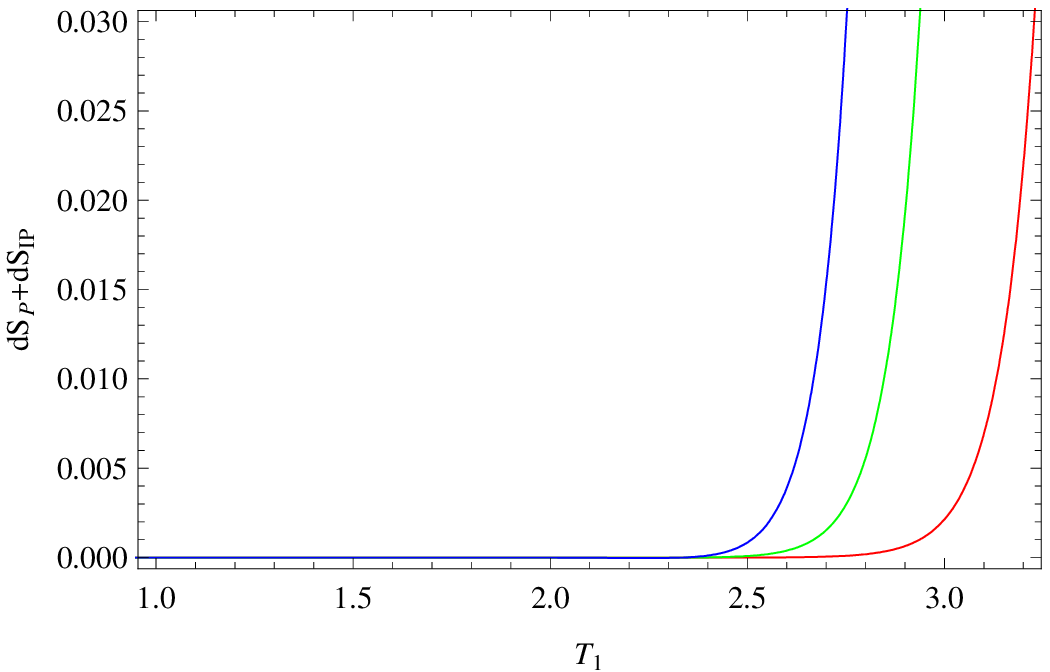}~~~~
\includegraphics[height=2in]{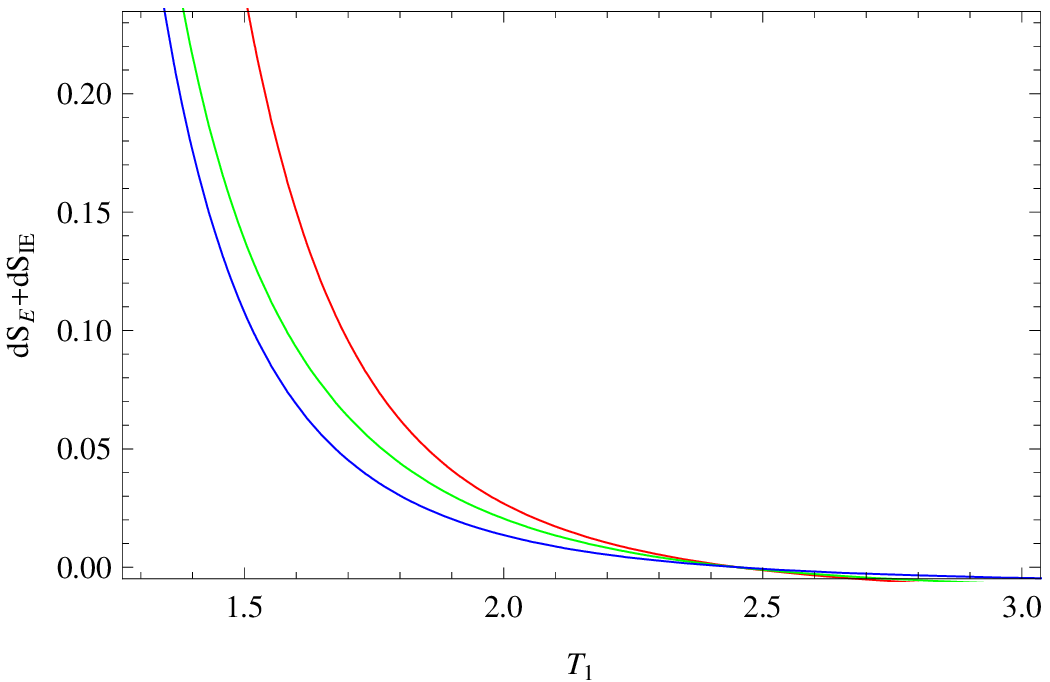}\\
\vspace{1mm} ~~~~~~~Fig.15~~~~~~~~~~~~~~~~~~~~~~~~~~~~~~~~~~~~
~~~~~~~~~~~~~~~~~~~~~~~~~~~~~~~~~~~~~Fig.16~~~\\

\vspace{6mm} Figs. 13, 14, 15 and 16 show the time derivatives of
the total entropy for Hubble horizon $R_{H}$, apparent horizon
$R_{A}$, particle horizon $R_{P}$ and event horizon $R_{E}$
respectively \textbf{without using first law of thermodynamics} in
the \emph{logamediate} scenario. The red, green and blue lines
represent the $dS_{X}+dS_{IX}$ for $k=-1,~1$ and $0$
respectively. We have chosen $~\xi=0.3$.\\

 \vspace{3mm}

 \end{figure}

From figures 13 to 16, we see that the total entropy increases
only in the case of particle horizon. In other words, the GSL
holds for all horizons in the Logamediate scenario. The maximum
rate of increase in entropy appears in the case of spatially flat
Universe while a hyperbolic curved Universe has the lowest rate of
increase in entropy. Also the behavior of the violation of GSL for
the apparent and future event horizon is quite similar. In
figs.14, the total change in entropy goes to zero in a finite time
while in fig.16, this rate remains constant but never tends to
zero.

\section{\normalsize\bf{GSL in Intermediate Scenario}}

We obtain the radii of Hubble, apparent, particle and event
horizons in intermediate expansion as

\begin{equation}
\begin{array}{c}
R_{H}=\frac{T_1^{1-\beta}}{B\beta}~;~~R_{A}=\frac{1}{\sqrt{ke^{-2BT_1^{\beta}}+B^{2}\beta^{2}T_1^{2(\beta-1)}}}~;\\\\
R_{P}=\frac{B^{-\frac{1}{\beta}}e^{BT_1^{\beta}}\left(\Gamma\left[\frac{1}{\beta}\right]-\Gamma\left[\frac{1}{\beta},BT_1^{\beta}\right]\right)}{\beta}~;
~~R_{E}=\frac{B^{-\frac{1}{\beta}}e^{BT_1^{\beta}}\Gamma\left[\frac{1}{\beta},BT_1^{\beta}\right]}{\beta} \\
\end{array}
\end{equation}

Using (17), (19) and (37) we can get the time derivatives of the
total entropies using as well as
without using the first law of thermodynamics.\\

\subsection{\bf GSL in the \emph{intermediate }scenario using first
law}

In this subsection we consider the GSL in the \emph{intermediate}
scenario. Using the first law the time derivatives of the total
entropies are \\

\begin{itemize}
    \item For Hubble horizon
    \begin{equation}
    \dot{S}_{H}+\dot{S}_{IH}=\frac{e^{-2BT_1^{\beta}} T_1^{-3\beta}(\beta-1)\left(-kT_1^{2}
    +Be^{2BT_1^{\beta}}T_1^{\beta}(\beta-\xi)\beta\right)}{B^{3}\beta^{3}GT_{H}}
    \end{equation}
    \item For apparent horizon
    \begin{equation}
    \dot{S}_{A}+\dot{S}_{IA}=\frac{B\beta e^{BT_1^{\beta}}T_1^{\beta}\left(kT_1^{2}-Be^{2BT_1^{\beta}}T_1^{\beta}
    (\beta-1)\beta\right)\left(kT_1^{2}-Be^{2BT_1^{\beta}}T_1^{\beta}
    (\beta-\xi)\beta\right)}{GT_{A}\left(kT_1^{2}+B^{2}e^{2BT_1^{\beta}}T_1^{2\beta}\beta^{2}\right)^{5/2}}
    \end{equation}
    \item For particle horizon
      \begin{equation}
      \dot{S}_{P}+\dot{S}_{IP}=\frac{B^{-\frac{3}{\beta}}\left(\Gamma\left[\frac{1}{\beta}\right]
      -\Gamma\left[\frac{1}{\beta},BT_1^{\beta}\right]\right)^{2}
      \left(kT_1^{2}-Be^{2BT_1^{\beta}}T_1^{\beta}(\beta-\xi)\beta\right)\left(B^{\frac{1}{\beta}}t+Be^{BT_1^{\beta}}
      T_1^{\beta}\left(\Gamma\left[\frac{1}{\beta}\right]
      -\Gamma\left[\frac{1}{\beta},BT_1^{\beta}\right]\right)\right)}{T_1^{3}\beta^{2}GT_{P}}\\
      \end{equation}
      \item   For event horizon
  \begin{equation}
   \dot{S}_{E}+\dot{S}_{IE}=\frac{B^{-\frac{3}{\beta}}\left(\Gamma\left[\frac{1}{\beta},BT_1^{\beta}\right]\right)^{2}
      \left(kT_1^{2}-Be^{2BT_1^{\beta}}T_1^{\beta}(\beta-\xi)\beta\right)\left(-B^{\frac{1}{\beta}}t+Be^{BT_1^{\beta}}
      T_1^{\beta}\Gamma\left[\frac{1}{\beta},BT_1^{\beta}\right]\right)}{T_1^{3}\beta^{2}GT_{E}}
    \end{equation}
\end{itemize}

\begin{figure}
 \includegraphics[height=2in]{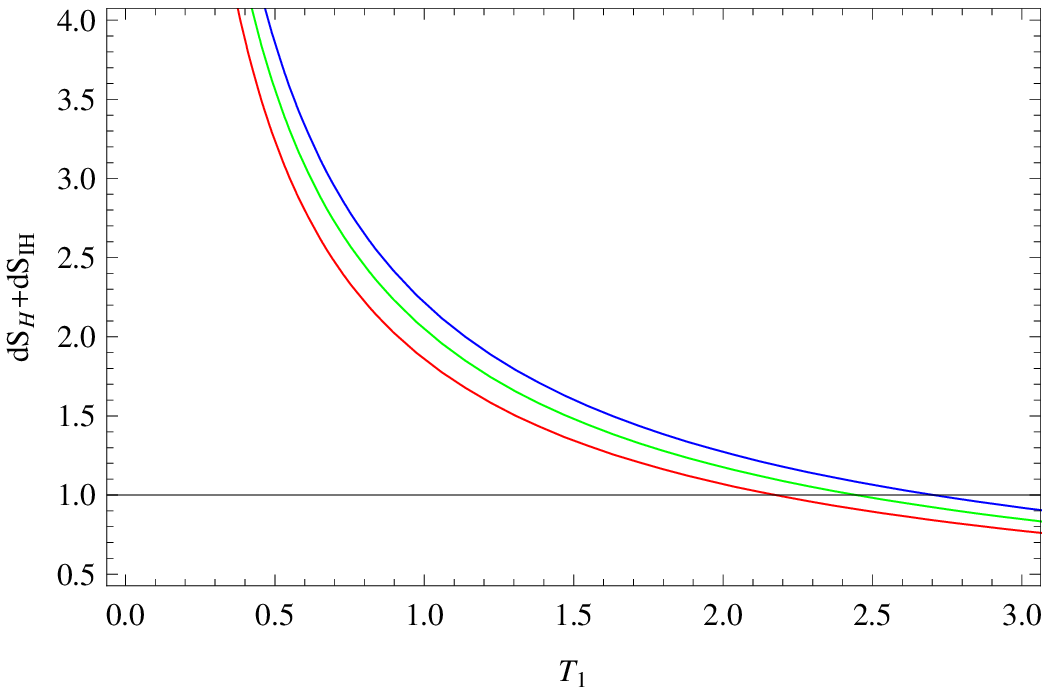}~~~~
\includegraphics[height=2in]{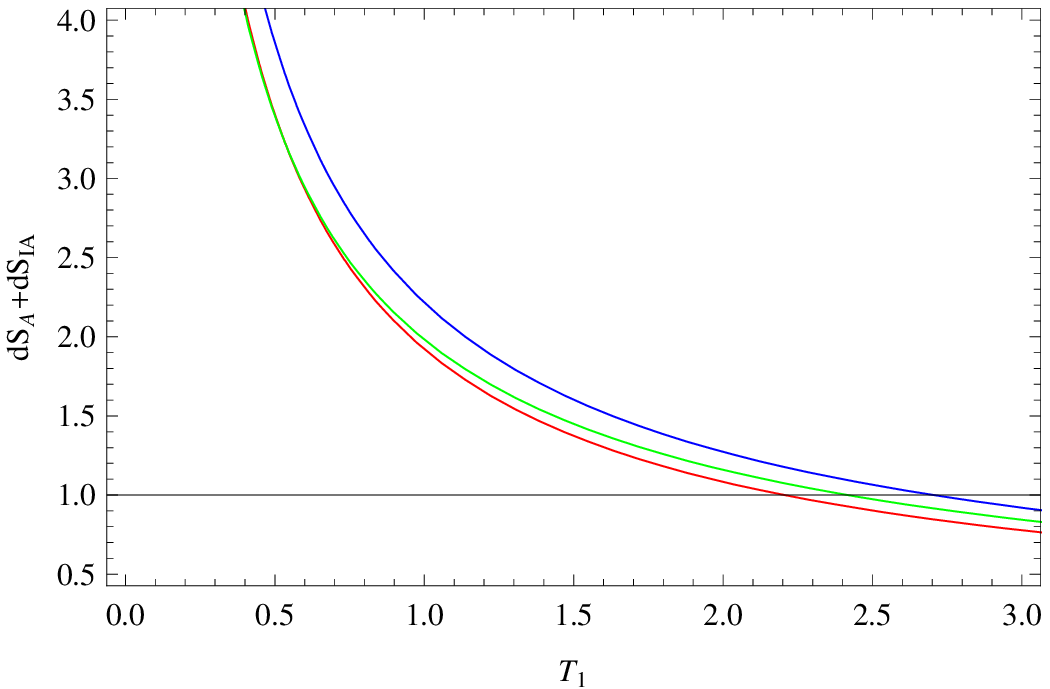}\\
\vspace{1mm} ~~~~~~~Fig.17~~~~~~~~~~~~~~~~~~~~~~~~~~~~~~
~~~~~~~~~~~~~~~~~~~~~~~~~~~~~~~~~~~~~~~~~~~Fig.18~~~\\

\vspace{6mm}
\includegraphics[height=2in]{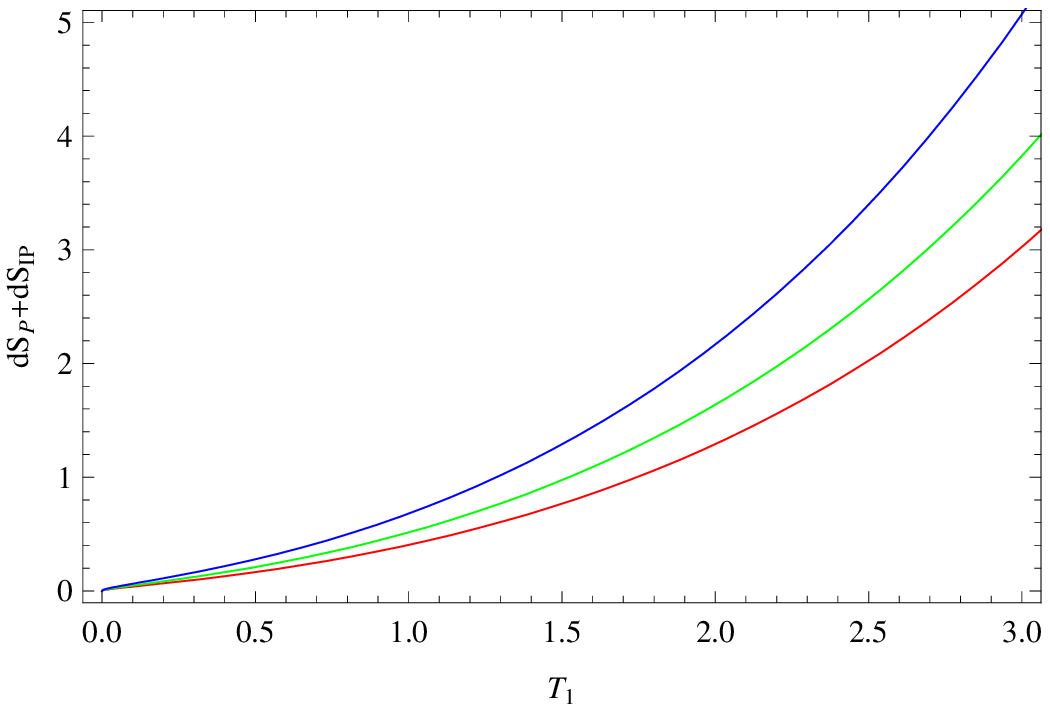}~~~~
\includegraphics[height=2in]{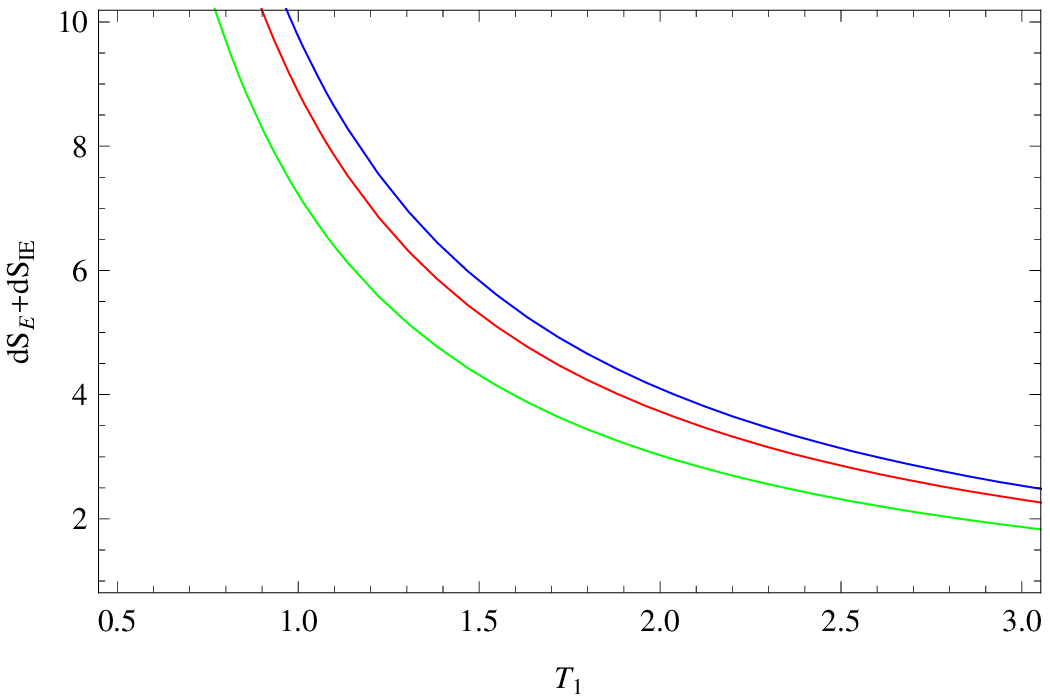}\\
\vspace{1mm} ~~~~~~~Fig.19~~~~~~~~~~~~~~~~~~~~~~~~~~~~~~
~~~~~~~~~~~~~~~~~~~~~~~~~~~~~~~~~~~~~~~~~~~Fig.20~~~\\

\vspace{6mm} Figs. 17, 18, 19 and 20 show the time derivatives of
the total entropy for Hubble horizon $R_{H}$, apparent horizon
$R_{A}$, particle horizon $R_{P}$ and event horizon $R_{E}$
respectively \textbf{using first law of thermodynamics} in the
\emph{intermediate} scenario. The red, green and blue lines
represent the $dS_{X}+dS_{IX}$ for $k=-1,~1$ and $0$
respectively. We have chosen $~\xi=0.3$.\\

 \vspace{6mm}

 \end{figure}

From figures 17 to 20, we see that the total entropy increases
only in the case of particle horizon. In other words, the GSL
holds for all horizons in the intermediate scenario. The maximum
rate of increase in entropy appears in the case of spatially flat
Universe while a hyperbolic curved Universe has the lowest rate of
increase in entropy. Also the behavior of the violation of GSL for
the apparent and future event horizon is quite similar. In fig.18,
the total change in entropy goes to zero in a finite time while in
fig.20, this rate remains constant but never tends to zero.

\subsection{\bf GSL in the \emph{intermediate} scenario without using
first law}

The time derivatives of the total entropies are also calculated
without using the first law of thermodynamics in the
\emph{intermediate} scenario as follows:\\

 \begin{figure}
 \includegraphics[height=2in]{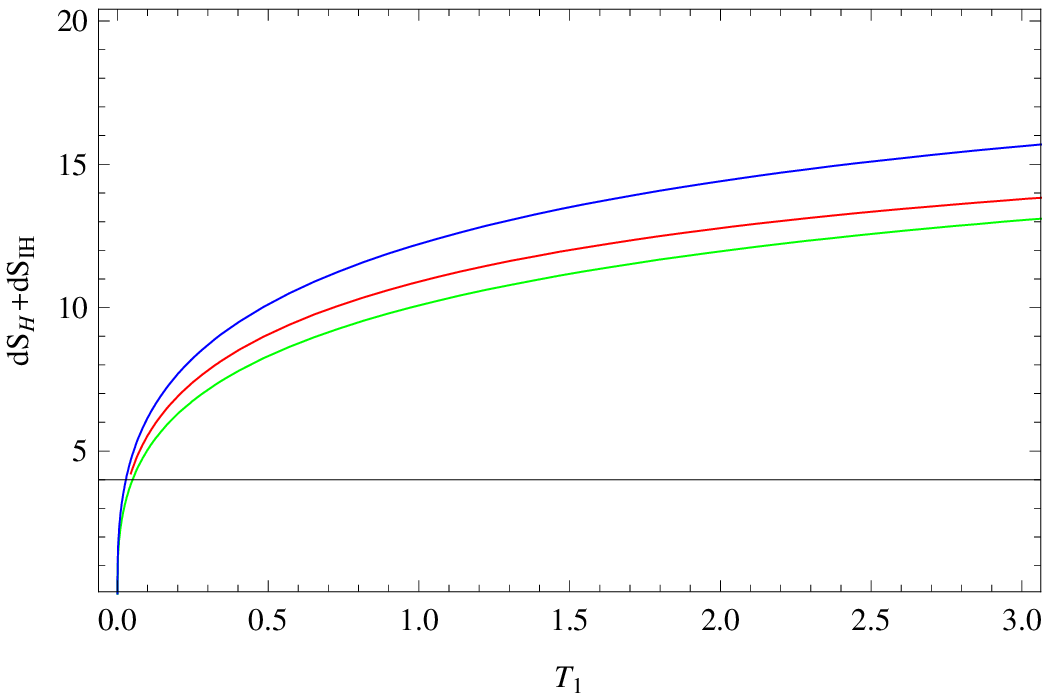}~~~~
\includegraphics[height=2in]{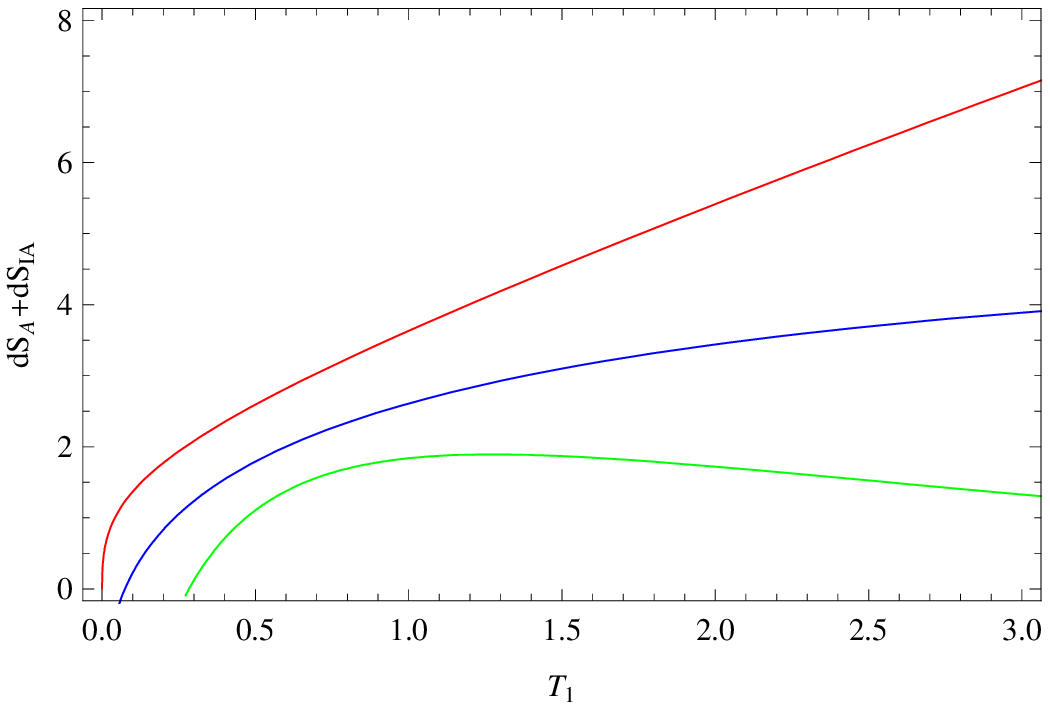}\\
\vspace{1mm} ~~~~~~~Fig.21~~~~~~~~~~~~~~~~~~~~~~~~~~~~~~~~~
~~~~~~~~~~~~~~~~~~~~~~~~~~~~~~~~~~~~~~~~Fig.22~~~\\

\vspace{6mm}
\includegraphics[height=2in]{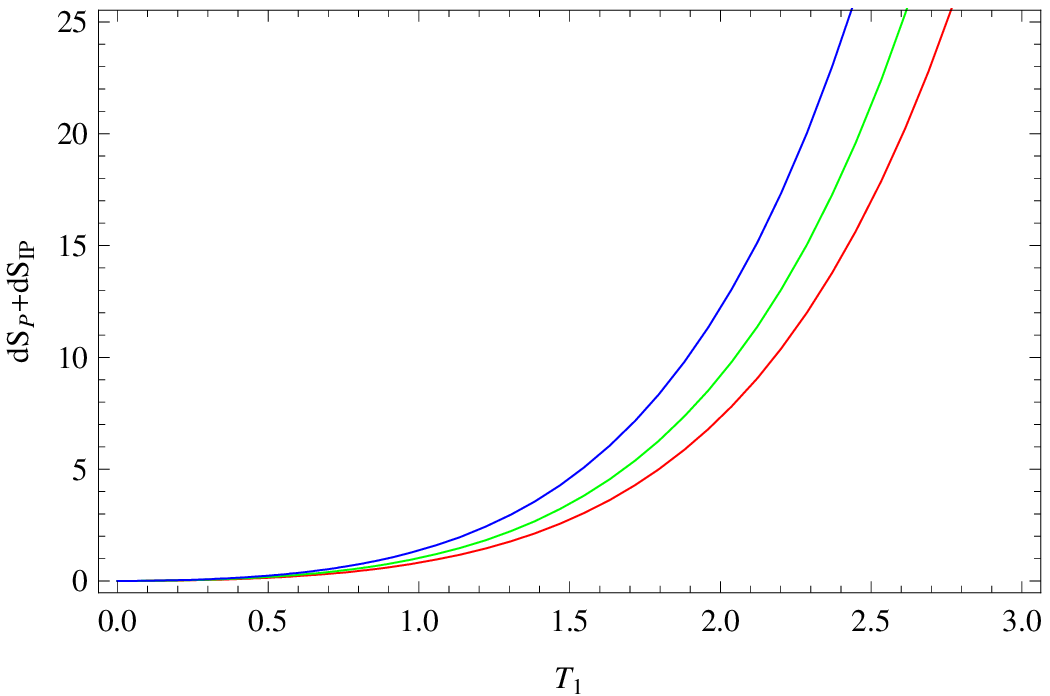}~~~~
\includegraphics[height=2in]{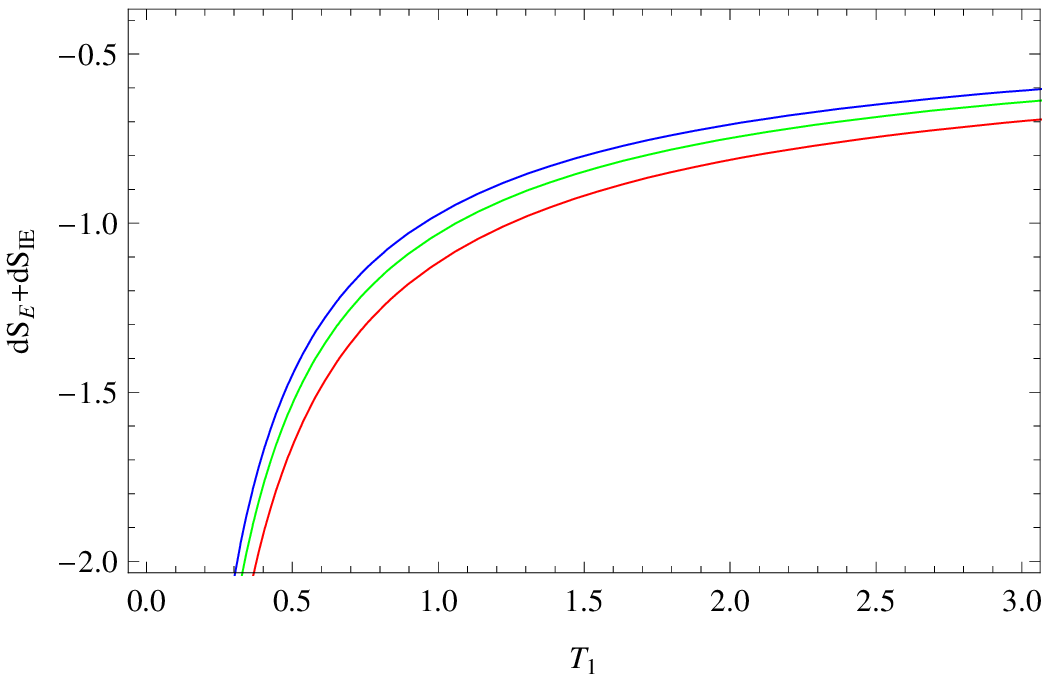}\\
\vspace{1mm} ~~~~~~~Fig.23~~~~~~~~~~~~~~~~~~~~~~~~~~~~~~
~~~~~~~~~~~~~~~~~~~~~~~~~~~~~~~~~~~~~~~~~~~Fig.24~~~\\

\vspace{6mm} Figs. 21, 22, 23 and 24 show the time derivatives of
the total entropy for Hubble horizon $R_{H}$, apparent horizon
$R_{A}$, particle horizon $R_{P}$ and event horizon $R_{E}$
respectively \textbf{without using first law of thermodynamics} in
the \emph{intermediate} scenario. The red, green and blue lines
represent the $dS_{X}+dS_{IX}$ for $k=-1,~1$ and $0$
respectively. We have chosen $~\xi=0.3$.\\

\vspace{6mm}

\end{figure}

\begin{itemize}
    \item For Hubble horizon
    \begin{equation}
    \dot{S}_{H}+\dot{S}_{IH}=\frac{2\pi e^{-2BT_1^{\beta}}
    T_1^{1-4\beta}\left[Be^{2BT_1^{\beta}}T_1^{\beta}\beta\left(B\beta T_1^{\beta}(1-\xi)+(\beta-1)(\beta-\xi) \right)+kT_1^{2}(1-\beta(1+BT_1^{\beta}))
    \right]}{B^{4}\beta^{4}G}
    \end{equation}
    \item For apparent horizon
    \begin{equation}
    \dot{S}_{A}+\dot{S}_{IA}=\frac{2\pi\beta Be^{2BT_1^{\beta}}T_1^{1+\beta}\left[ \left(kT_1^{2}-Be^{2BT_1^{\beta}}T_1^{\beta}
    (\beta-1)\beta\right)^{2} +(\xi-1)B^{2}\beta^{2}T_1^{2\beta}e^{4BT_1^{\beta}}(1-\beta(1+BT_1^{\beta})) \right]}{G\left(kT_1^{2}+B^{2}\beta^{2}e^{2BT_1^{\beta}}T_1^{2\beta}\right)^{3}}
    \end{equation}
    \item For particle horizon
       \begin{eqnarray*}
      \dot{S}_{P}+\dot{S}_{IP}=\frac{2\pi B^{-\frac{1}{\beta}}e^{BT_1^{\beta}}\left(\Gamma\left[\frac{1}{\beta}\right]
      -\Gamma\left[\frac{1}{\beta},BT_1^{\beta}\right]\right)}{\beta^{3}G}\times
       \left[\beta^{2}+B^{1-\frac{1}{\beta}}\beta^{2}e^{BT_1^{\beta}}T_1^{\beta-1}
       \left(\Gamma\left[\frac{1}{\beta}\right]-\Gamma\left[\frac{1}{\beta},BT_1^{\beta}\right]\right)\right.
    \end{eqnarray*}
    \begin{equation}
~~~~~~~~~~~~~~~~~\left.+B^{-\frac{2}{\beta}}
\left(k-Be^{2BT_1^{\beta}}T_1^{\beta-2}(\beta-\xi)\beta\left(\Gamma
       \left[\frac{1}{\beta}\right]-\Gamma\left[\frac{1}{\beta},BT_1^{\beta}\right]\right)^{2}\right)\right]
    \end{equation}
    \item For event horizon
    \begin{eqnarray*}
    \dot{S}_{E}+\dot{S}_{IE}=
    \frac{2\pi
    B^{-\frac{3}{\beta}}e^{BT_1^{\beta}}\Gamma\left[\frac{1}{\beta},BT_1^{\beta}\right]}{T_1^{2}\beta^{3}G}\times
    \left[\beta Be^{BT_1^{\beta}}T_1^{\beta}\left(\beta B^{\frac{1}{\beta}}t+(\beta-\xi)e^{BT_1^{\beta}}
    \Gamma\left[\frac{1}{\beta},BT_1^{\beta}\right]\right) \right.
    \end{eqnarray*}
 \begin{equation}
~~~~~~~~~~~~~~~~~~~~ \left.
-T_1^{2}\left(\beta^{2}B^{\frac{2}{\beta}}+k
    \left(\Gamma\left[\frac{1}{\beta},BT_1^{\beta}\right]\right)^{2}  \right) \right]
    \end{equation}
\end{itemize}

From figures 21 to 24, we see that the total entropy increases in
the all horizons. The the GSL holds only for Hubble, apparent and
particle horizons but does not hold on event horizon in the
intermediate scenario.

\section{GSL in the Power Law form of the expansion}

We consider the power law form of the scale factor as

\begin{equation}
a=a_{0}T_1^{m}
\end{equation}
 where $m>0$. This form of scale factor has been used earlier by \cite{powerlaw}. For acceleration of the universe, $m>1$. So in this case the particle horizon does not
 exist.  Particle horizon exists only for $0<m<1$, i.e., in decelerating phase. The radius of particle horizon is

\begin{equation}
R_{P}=\frac{T_1}{1-m}~~with~~0<m<1
\end{equation}

The radii of Hubble, apparent and event horizons are respectively

\begin{equation}
R_{H}=\frac{T_1}{m}~,~~R_{A}=\left(\frac{n^{2}}{T_1^{2}}
+\frac{k}{a_{0}^{2}T_1^{2m}} \right)^{-\frac{1}{2}}~~and
~~R_{E}=\frac{T_1}{m-1}~~with~~m>1
\end{equation}

\subsection{\bf GSL in the power law form using first
law}

In this subsection we consider the GSL in the power law form. Using
the first law the time derivatives of the total
entropies are \\

\begin{itemize}
    \item For Hubble horizon
    \begin{equation}
    \dot{S}_{H}+\dot{S}_{IH}=\frac{kT_1^{2-2m}+a_{0}^{2}m\xi }{a_{0}^{2}m^{3}GT }
    \end{equation}
    \item For apparent horizon
    \begin{equation}
    \dot{S}_{A}+\dot{S}_{IA}=\frac{a_{0}^{2}mT_1^{m} (kT_1^{2}+a_{0}^{2}mT_1^{2m} )(kT_1^{2}+a_{0}^{2}m\xi T_1^{2m} )}{GT (kT_1^{2}+a_{0}^{2}m^{2}T_1^{2m})^{\frac{5}{2}} }
    \end{equation}
    \item For particle horizon
      \begin{equation}
      \dot{S}_{P}+\dot{S}_{IP}=\frac{T_1^{-2m}(kT_1^{2}+a_{0}^{2}m\xi T_1^{2m})}{a_{0}^{2}GT(1-m)^{3} }
      \end{equation}
      \item   For event horizon
  \begin{equation}
   \dot{S}_{E}+\dot{S}_{IE}=\frac{T_1^{-2m}(kT_1^{2}+a_{0}^{2}m\xi T_1^{2m})}{a_{0}^{2}GT(m-1)^{3} }
    \end{equation}
\end{itemize}

Figures 25 to 28 show the nature of the variation of total entropy
in the universe where the scale factor is evolving in power law
form in the framework of FAC using the first law of
thermodynamics. For $k=0$, the variation of total entropy does not
change with the evolution of the universe. However, it stays at
positive level. For $k=1$, the GSL of thermodynamics holds for all
types of horizons. However, for the universe enveloped by Hubble
horizon, we find that the time derivative of the total entropy is
negative for $k=-1$. This means the breaking down of the GSL.  The
same holds for particle horizon. In the case of apparent horizon
enclosing the open universe, the GSL breaks down in the early
stage of the universe. However, at later stages, the GSL holds.
The same is true, when we consider the universe enveloped by the
event horizon.

\begin{figure}
 \includegraphics[height=2in]{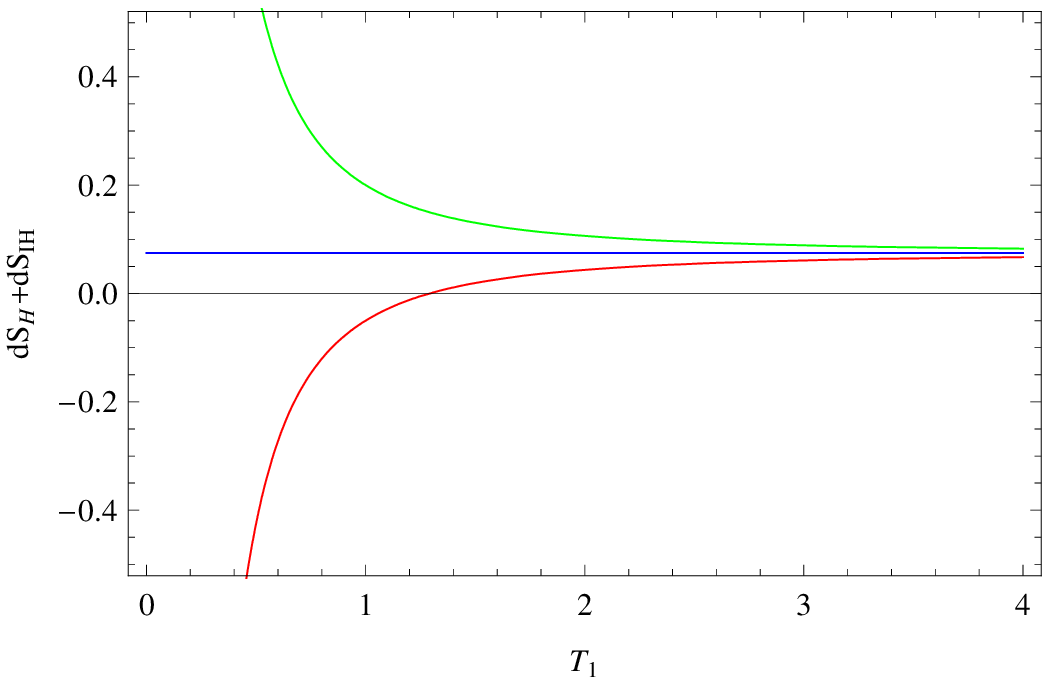}~~~~
\includegraphics[height=2in]{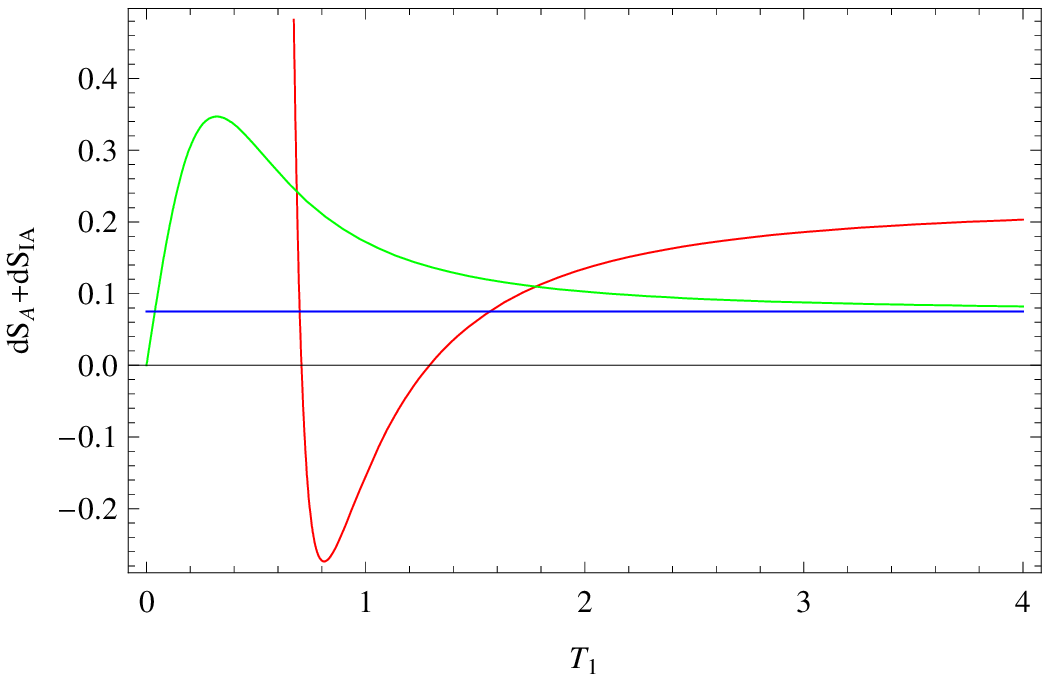}\\
\vspace{1mm} ~~~~~~~Fig.25~~~~~~~~~~~~~~~~~~~~~~~~~~~~~~~
~~~~~~~~~~~~~~~~~~~~~~~~~~~~~~~~~~~~~~~~~~Fig.26~~~\\

\vspace{6mm}
\includegraphics[height=2in]{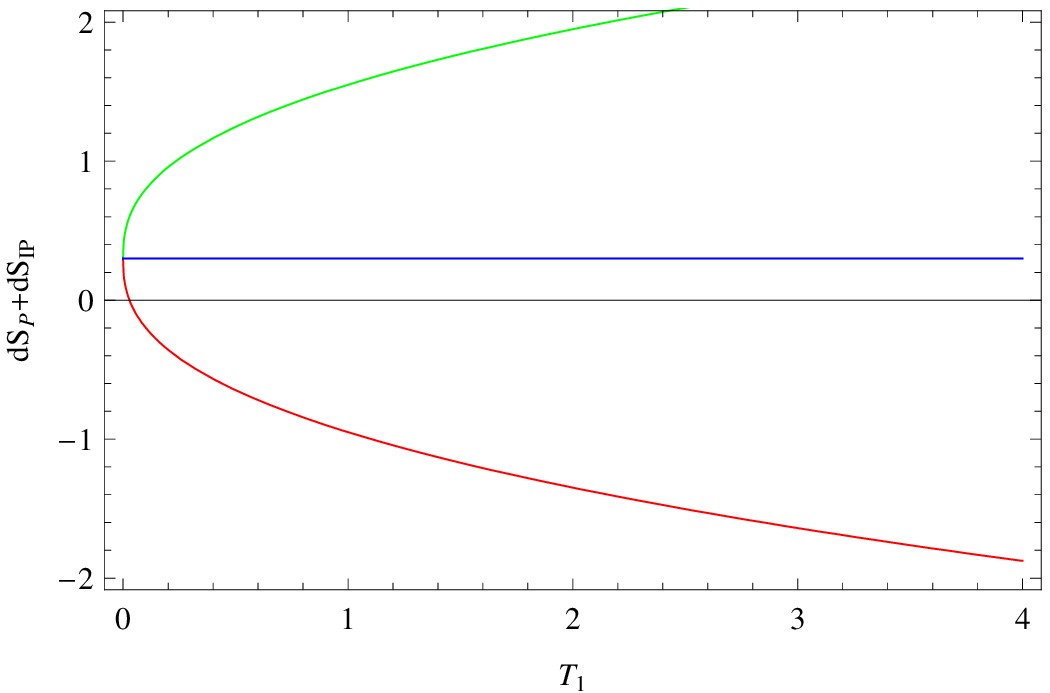}~~~~
\includegraphics[height=2in]{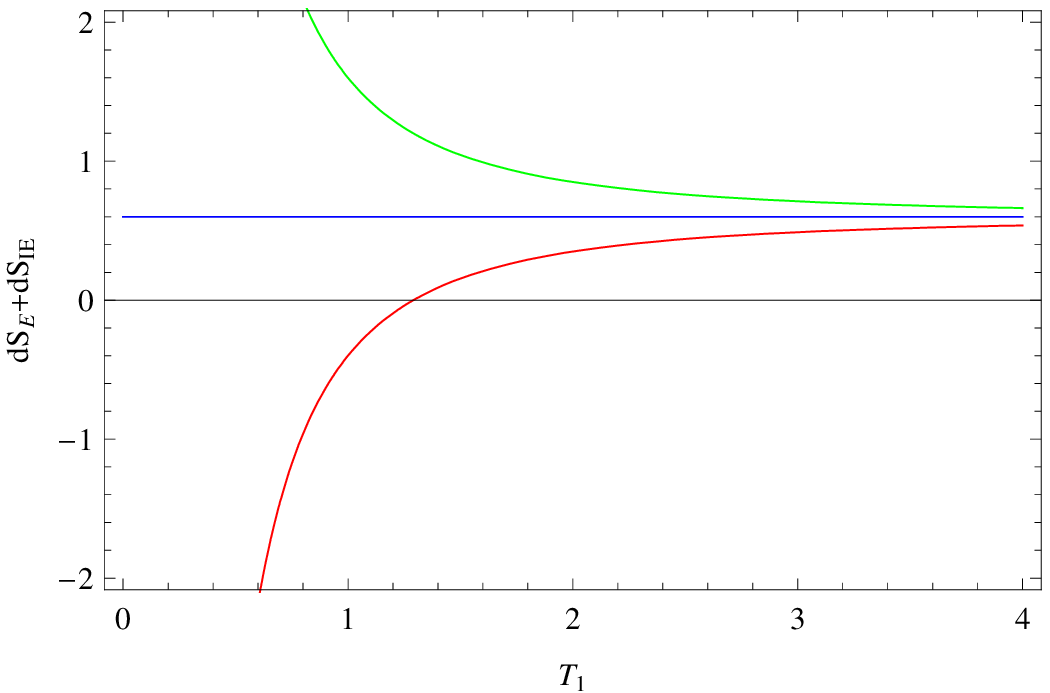}\\
\vspace{1mm} ~~~~~~~Fig.27~~~~~~~~~~~~~~~~~~~~~~~~~~~~~~
~~~~~~~~~~~~~~~~~~~~~~~~~~~~~~~~~~~~~~~~~~~Fig.28~~~\\

\vspace{6mm} Figs. 25, 26, 27 and 28 show the time derivatives of
the total entropy for Hubble horizon $R_{H}$, apparent horizon
$R_{A}$, particle horizon $R_{P}$ and event horizon $R_{E}$
respectively \textbf{using first law of thermodynamics} in the
power law form. The red, green and blue lines represent the
$dS_{X}+dS_{IX}$ for $k=-1,~1$ and $0$ respectively. We have
chosen $a_{0}=1,~~\xi=.3$. For particle horizon, we have chosen
$m=.5$ and for other horizons, $m=2$.\\

\vspace{6mm}

\end{figure}

\begin{figure}
 \includegraphics[height=2in]{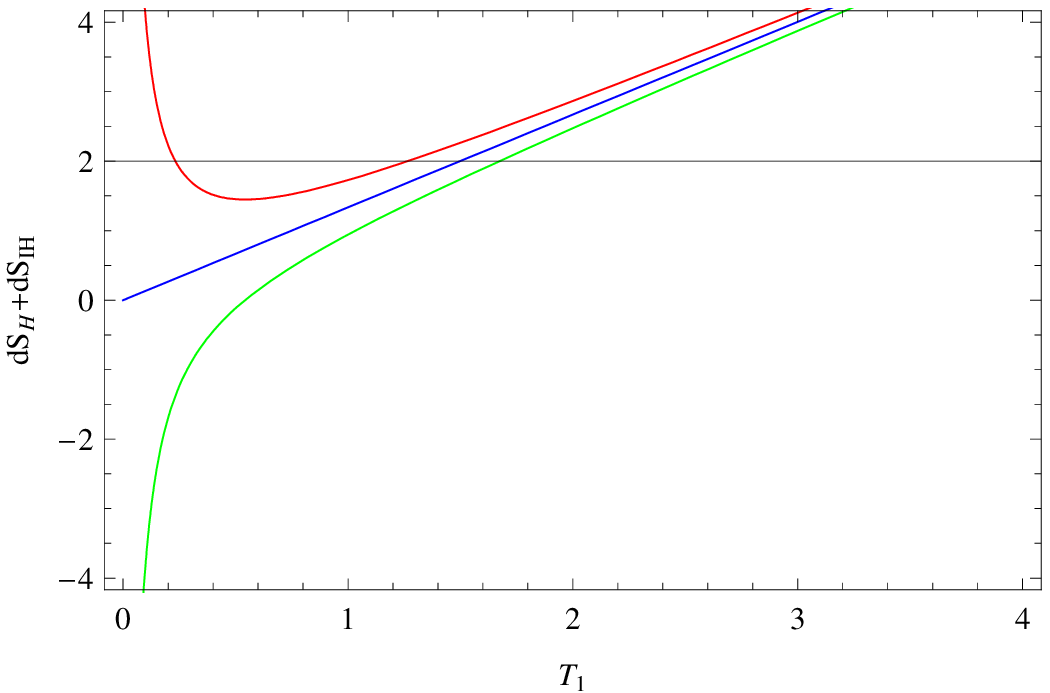}~~~~
\includegraphics[height=2in]{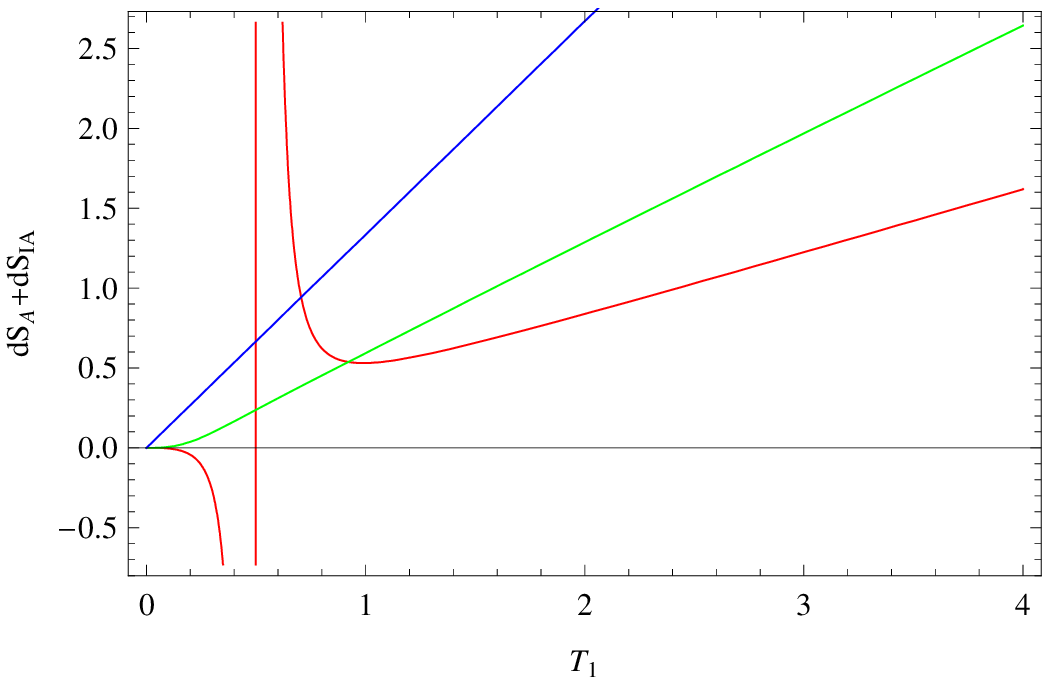}\\
\vspace{1mm} ~~~~~~~Fig.29~~~~~~~~~~~~~~~~~~~~~~~~~~~~~~~~~~~~~
~~~~~~~~~~~~~~~~~~~~~~~~~~~~~~~~~~~~Fig.30~~~\\

\vspace{6mm}
\includegraphics[height=2in]{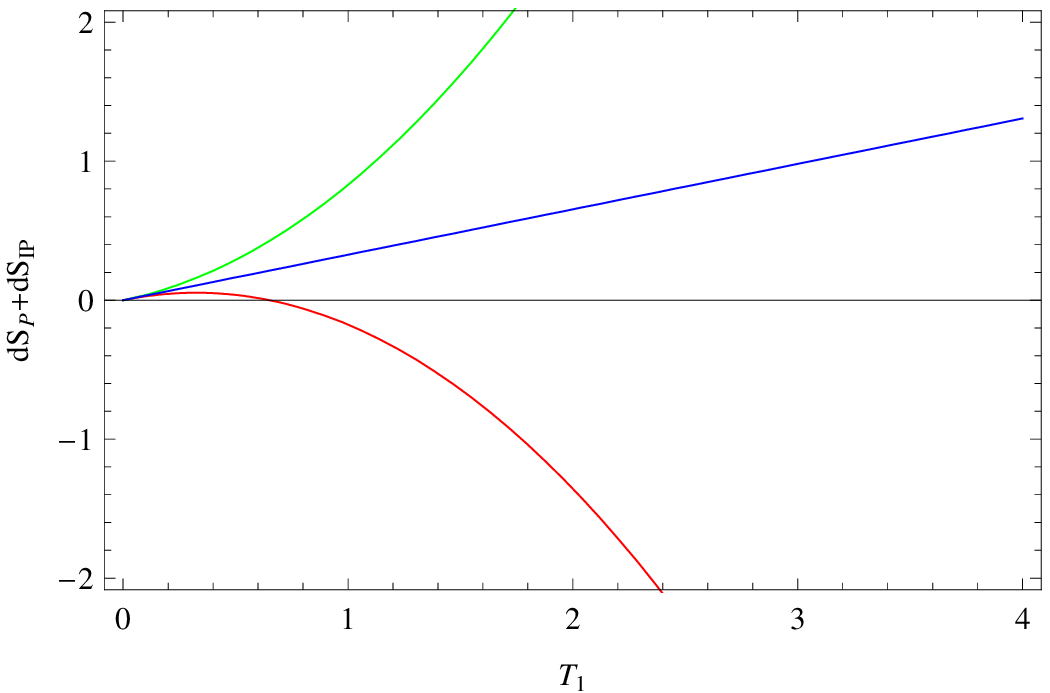}~~~~
\includegraphics[height=2in]{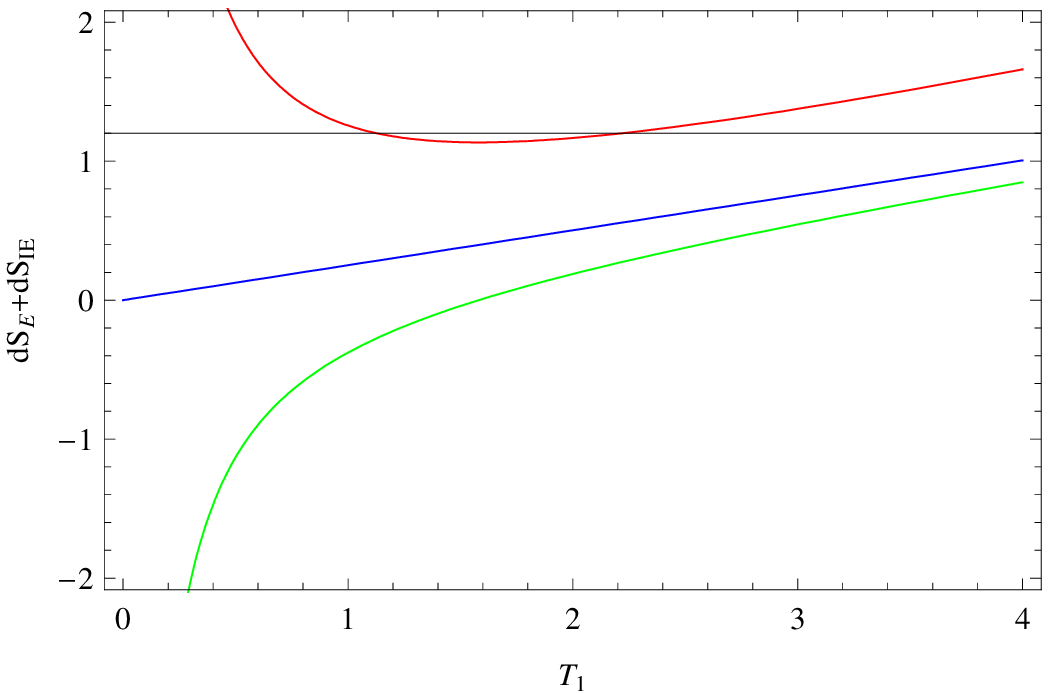}\\
\vspace{1mm} ~~~~~~~Fig.31~~~~~~~~~~~~~~~~~~~~~~~~~~~~~~~~~~~~~~~
~~~~~~~~~~~~~~~~~~~~~~~~~~~~~~~~~~Fig.32~~~\\

\vspace{6mm} Figs. 29, 30, 31 and 32 show the time derivatives of
the total entropy for Hubble horizon $R_{H}$, apparent horizon
$R_{A}$, particle horizon $R_{P}$ and event horizon $R_{E}$
respectively \textbf{without using first law of thermodynamics} in
the power law form. The red, green and blue lines represent the
$dS_{X}+dS_{IX}$ for $k=-1,~1$ and $0$ respectively. We have
chosen $a_{0}=1,~~\xi=.3$. For particle horizon, we have chosen
$m=.5$ and for other horizons, $m=2$.\\

\vspace{6mm}

\end{figure}

\subsection{\bf GSL in the power law form without using
first law}

The time derivatives of the total entropies are also calculated
without using the first law of thermodynamics in the
power law form  as follows:\\

\begin{itemize}
    \item For Hubble horizon
    \begin{equation}
    \dot{S}_{H}+\dot{S}_{IH}=\frac{2k(1-m)\pi T_1^{3-2m}+2a_{0}^{2}m\pi (m+\xi-m\xi)T_1 }{a_{0}^{2}m^{4}G}
    \end{equation}
    \item For apparent horizon
    \begin{equation}
    \dot{S}_{A}+\dot{S}_{IA}=\frac{2a_{0}^{2}m\pi T_1^{1+2m} (k^{2}T_1^{4}+2a_{0}^{2}kmT_1^{2+2m}-a_{0}^{4}m^{2}(m\xi-\xi-m)T_1^{4m} )}{G (kT_1^{2}+a_{0}^{2}m^{2}T_1^{2m})^{3} }
    \end{equation}
    \item For particle horizon
      \begin{equation}
      \dot{S}_{P}+\dot{S}_{IP}=\frac{2\pi T_1(kT_1^{2-2m}+a_{0}^{2}(m\xi-m+1))}{a_{0}^{2}G(1-m)^{3} }
      \end{equation}
      \item   For event horizon
  \begin{equation}
   \dot{S}_{E}+\dot{S}_{IE}=\frac{2\pi T_1(kT_1^{2-2m}+a_{0}^{2}(m\xi-m+1))}{a_{0}^{2}G(1-m)^{3} }
    \end{equation}
\end{itemize}

The evolutions of the time derivatives of the total entropy for
the four horizons are displayed in the figures 29 to 32. For
$k=-1$, the GSL always holds for event horizon and always breaks
down for particle horizon. However, breaks down at early stages
and holds at later stages for the universes enveloped by Hubble
and apparent horizon. For $k=1$ the GSL always holds for apparent
horizon and particle horizon and always breaks down for event
horizon. However, breaks down at early stages and holds at later
stages for the universe enveloped by the Hubble horizon. For $k=0$
the GSL always holds for apparent, particle and event horizons.
Here also the GSL breaks down at early stages and holds at later
stages for the universe enveloped by the Hubble horizon.\\

\section{Conclusion}
In the framework of Fractional Action Cosmology (FAC), we study
the generalized second law of thermodynamics for the Friedmann
Universe enclosed by a boundary. Use of the first law of
thermodynamics led to the satisfaction of the GSL of
thermodynamics in the emergent scenario irrespective of the type
of the horizon we consider. This is reflected in the figures 1 to
4, where the time derivatives of the total entropy stayed at the
positive level for Hubble, apparent, particle and event horizons.
In the case of the particle horizon, the time derivative of the
total entropy decayed with cosmic time. On the contrary, it
exhibited increasing pattern in the cases where the universe was
supposed to be enveloped by the other three horizons. The
behaviors of the total entropies did not show any variation with
respect to the curvature of the universe. When we considered the
entropy for all of the four horizons without taking the first law
of thermodynamics into account we observed the behaviors similar
to the cases of using the first law of thermodynamics. This is
reflected in the figures 4 to 8. In the logamediate scenario, the
time derivatives of the total entropy were found to be decreasing
function of cosmic time excepting the universe enveloped by the
particle horizon. This does not depend on whether we are using or
not using the first law of thermodynamics to study the GSL of
thermodynamics. It must be noted that in all of the cases
considered under logamediate scenario, the time derivative of the
total entropy remains positive. This indicates the validity of the
GSL of thermodynamics in the logamediate scenario within the
framework of FAC. Figures 9 to 16 reflect this behavior. Figures
17 to 20 repeat the similar behavior of the time derivative of the
total entropy when the first law is used to investigate the GSL of
thermodynamics in the intermediate scenario. However, when we
ignore the first law of thermodynamics, we find that the time
derivative of the total entropy fails to stay at positive level in
the universe enveloped by the event horizon (see figure 24).
However, figures 21 to 23 show an increasing nature of the
positive time derivative of the total entropy in the universes
enveloped by Hubble, apparent and particle horizon respectively
under intermediate scenario without using the first law of
thermodynamics in the framework of FAC.

Figures 25 to 28 show the nature of the time derivative of the
total entropy in the universe where the scale factor is evolving
in power law form in the framework of FAC using the first law of
thermodynamics. For $k=0$, that is, in the flat universe, the time
derivative of the total entropy does not change with the evolution
of the universe. However, it stays at positive level. For $k=1$,
the GSL of thermodynamics holds for all types of horizons.
However, for the universe enveloped by Hubble horizon, we find
that the time derivative of the total entropy is negative for
$k=-1$. This means the breaking down of the GSL.  The same holds
for particle horizon. In the case of apparent horizon enclosing
the open universe, the GSL breaks down in the early stage of the
universe. However, at later stages, the GSL holds under the
framework of FAC. The same is true, when we consider the universe
enveloped by the event horizon. Next we consider the power law
scenario without using the first law of thermodynamics. The
evolutions of the time derivatives of the total entropy for the
four horizons are displayed in the figures 29 to 32. For $k=-1$,
the GSL always holds for event horizon and always breaks down for
particle horizon. However, breaks down at early stages and holds
at later stages for the universes enveloped by Hubble and apparent
horizon. For $k=1$ the GSL always holds for apparent horizon and
particle horizon and always breaks down for event horizon.
However, breaks down at early stages and holds at later stages for
the universe enveloped by the Hubble horizon. For $k=0$ the GSL
always holds for apparent, particle and event horizons. Here also
the GSL breaks down at early stages and holds at later stages for
the universe enveloped by the Hubble horizon.

In summary, we investigated the validity of the generalized second
law of thermodynamics in the framework of Fractional Action
Cosmology (FAC). We applied this law to study the thermodynamics
for the Friedmann Universe. To enclose the Universe by a boundary,
we used the four well-known cosmic horizons as boundaries namely,
apparent horizon, future event horizon, Hubble horizon and
particle horizon. Using the two possible approaches, we
constructed the GSL using and without using the first law of
thermodynamics. To simplify our analysis, we preferred express the
law in the form of four different scale factors namely emergent,
logamediate, intermediate and power law. For Hubble, apparent and
particle horizons, the GSL holds for emergent and logamediate
expansions of the universe when we apply with and without using
first law. For intermediate scenario, the GSL is valid for Hubble,
apparent, particle horizons when we apply with and without first
law. Also for intermediate scenario, the GSL is valid for event
horizon when we apply first law but it breaks down without using
first law. But for power law expansion, the GSL may be valid for
some cases and breaks down otherwise.

\end{document}